\documentclass[aps,epsf,showpacs,superscriptaddress]{revtex4}
\usepackage{amsmath}
\usepackage{epsfig}

\newcommand{\T}{{\mathcal{T}}}
\newcommand{\be}{\begin{equation}}
\newcommand{\ee}{\end{equation}}

\def\proba{{\rm I\kern -.18em P}}

\newcommand{\I}{{\rm{i}}}
\newcommand{\D}{{\rm{d}}}

\newcommand{\diff}{{\rm d}}

\begin{document}

%
\title{Semiclassical form factor for spectral and matrix element 
       fluctuations of multidimensional chaotic systems}

\author{Marko Turek}
\email{marko.turek@physik.uni-regensburg.de}
\affiliation{Institut f{\"u}r Theoretische Physik, Universit{\"a}t Regensburg,
             D-93040 Regensburg, Germany}
\author{Dominique Spehner}
\affiliation{Fachbereich Physik, Universit{\"a}t Duisburg--Essen, D-45117 Essen, 
             Germany}
\author{Sebastian M{\"u}ller}
\affiliation{Fachbereich Physik, Universit{\"a}t Duisburg--Essen, D-45117 Essen, 
             Germany}
\author{Klaus Richter}
\affiliation{Institut f{\"u}r Theoretische Physik, Universit{\"a}t Regensburg,
             D-93040 Regensburg, Germany}

\date{\today}

\begin{abstract}

We present a semiclassical calculation of the generalized
form factor $K_{ab}(\tau)$ which characterizes the fluctuations
of matrix elements of the operators $\hat{a}$ and
$\hat{b}$ in the eigenbasis of the Hamiltonian of a chaotic system. 
Our approach is
based on some recently developed techniques for the spectral
form factor of systems with hyperbolic and ergodic underlying classical dynamics
and $f=2$ degrees of freedom,
 that allow us to go beyond the diagonal approximation.
First we extend these techniques to
systems with $f>2$. Then we use
these results to calculate $K_{ab}(\tau)$. We show that the
dependence on the rescaled time $\tau$ (time in units of the Heisenberg time)
is universal for both the spectral and the generalized form factor. 
Furthermore, we derive a relation between $K_{ab}(\tau)$ and the
classical time--correlation function of the Weyl symbols of $\hat{a}$ and $\hat{b}$.
\end{abstract}


\maketitle


%

\section{Introduction}
\label{sec:introduction}

\subsection{Overview}
\label{subsec-overview}

In a number of recent works the quantum spectral statistics
of closed chaotic  systems was investigated in the
semiclassical limit. According to a conjecture by 
Bohigas, Giannoni and Schmit~\cite{BGS84} (BGS),
the fluctuations of the energy levels are
system--independent and coincide
with the predictions of random--matrix theory (RMT)
if the system has 
a chaotic underlying classical dynamics.
Numerical and experimental investigations carried out 
on a great variety of systems support the 
BGS conjecture~\cite{Haake,Stoeckmann}. 

In the semiclassical limit, Gutzwiller's trace formula~\cite{Gutzwiller90}
provides a suitable starting point for the calculation
of spectral correlation functions. It relates
the quantum mechanical density of states to a sum over
classical periodic orbits which are characterized by an amplitude and
a phase that is obtained from the action of the
orbit. A prominent example of a quantum correlation
function is the two--point
energy--energy correlation function
and its Fourier transform, the spectral form factor $K(\tau)$.
In this case a semiclassical analysis faces the problem
of evaluating a double sum over periodic orbits
which requires an appropriate quantitative
treatment of classical action correlations.
Averaging the form factor of a given system
over an energy window that is large compared to
the mean level spacing implies that only pairs of
orbits with a small action difference of the
order of $\hbar$ yield significant contributions.
The leading contribution given by the terms
with vanishing action difference is obtained within
the diagonal approximation~\cite{Berry85}. This approximation
accounts for all orbit pairs where an orbit is
associated to itself or to its time--reversed partner
if time--reversal symmetry is present. 

Only recently a method for a systematic 
inclusion of certain orbit pairs with small but nonzero
action differences was developed for systems with time--reversal
symmetry~\cite{Sieber01,Sieber02}.
This approach, originally
formulated in the configuration space, was first applied
to the next-to-leading order correction of the
spectral form factor of a uniformly
hyperbolic system 
showing agreement with the RMT predictions.
In subsequent works, extensions to a phase--space
formulation applicable to
nonuniformly hyperbolic systems~\cite{Spehner03,Turek03}
and to higher--order
corrections~\cite{Mueller04} were proposed.
However, all these previous considerations were
restricted to systems with $f=2$ degrees of freedom, e.g.,
two--dimensional billiards, while the more general RMT
conjecture is independent of  $f$. In this work, we present
a theory which applies to hyperbolic and ergodic Hamiltonian systems with 
an arbitrary number $f$ of degrees of freedom ($f \geq 2$).
These systems are characterized by a
set of system--specific time scales, namely the
$(f-1)$ positive Lyapunov exponents. However, we will prove
that going beyond the diagonal approximation
the final result for the spectral
form factor is independent of $f$
and coincides in a universal way with the RMT predictions.

From an experimental point of view it is desirable
to furthermore develop a theory that describes 
not only statistical properties of the energy
spectrum but also of quantum mechanical matrix elements, 
as entering, for example, in cross sections. 
The fluctuations of the diagonal matrix elements
of the operators $\hat{a}$ and $\hat{b}$ in the eigenbasis
of the Hamiltonian can be described by
a generalized form factor~\cite{Eckhardt95} $K_{ab}(\tau)$.

Similarly to the spectral form factor,
one expects that $K_{ab} (\tau)$ also shows 
universal features as $\hbar \to 0$
and depends only on averaged classical quantities, 
like the averages over the constant--energy surface and the time correlation function of the 
Weyl symbols of $\hat{a}$ and $\hat{b}$.
Analytical results exclusively based on the diagonal approximation
have to some extent confirmed this 
statement~\cite{Eckhardt95,EFKAMM95,Eckhardt97,Eckhardt00}.
In this work, we generalize  these results beyond the diagonal 
approximation. 

In the following two subsections
we briefly recall the semiclassical theory 
for the form factor based
on Gutzwiller's trace formula.
In section~\ref{sec:higher_dimensional}, 
we discuss the diagonal approximation and the origin of the off--diagonal corrections 
in the special case of  the spectral form factor. 
We then show how the
results for two--dimensional systems
can be extended to higher--dimensional ones, leading
once more to universality and agreement with RMT predictions
in the semiclassical limit.
The matrix--element fluctuations described by the generalized
form factor  $K_{ab}(T)$ are then studied 
in Section~\ref{sec:matrix-element_fluct}.
The leading--order 
(diagonal approximation) and next-to-leading-order terms
are determined. 

\subsection{Generalized form factor: definitions and main results}
\label{subsec-gen_K}

We introduce the  weighted density of states
\begin{equation}
\label{gen_density}
 d_a(E) \equiv {\rm tr}  \left( \hat{a}\, \delta ( E - \hat{H} ) \right) =
  \sum\limits_n \langle n | \hat{a} | n \rangle\,
 \delta (E - E_n)\,
\end{equation}
for a quantum observable $\hat{a}$. This is in generalization
of the spectral density of states where $\hat{a}$ is given by the
identity operator, i.e., $\hat{a} = \hat{1}$.
In Eq.~(\ref{gen_density}),
$|n \rangle$ are the eigenstates and
$E_n$ the corresponding eigenenergies
of the Hamiltonian $\hat{H}$ of the system.
The two--point correlation function
\begin{widetext}
\begin{equation} \label{eq-correl_function}
R_{ab} (\epsilon) 
 \equiv \frac{1}{\left\langle d(E) \right\rangle_{\Delta E}^2}
  \left( \left\langle d_a \left(E - \frac{\epsilon}{2} \right) 
   d_b \left( E + \frac{\epsilon}{2} \right) \right\rangle_{\Delta E}  
   -  \biggl\langle d_a (E  ) \biggr\rangle_{\Delta E} 
      \biggl\langle d_b (E  ) \biggr\rangle_{\Delta E} \right)
\end{equation}
\end{widetext}
describes correlations between the diagonal matrix
elements of $\hat{a}$ and $\hat{b}$ in the eigenbasis $\{ | n \rangle \}$. 
In Eq.~(\ref{eq-correl_function}), 
$\left\langle d(E) \right\rangle_{\Delta E}$ is the
mean density of states, given by 
$\left\langle d(E) \right\rangle_{\Delta E} = \Omega /(2 \pi \hbar)^{f}$,
where $f$ is the number 
of degrees of freedom of the system and
$\Omega = \int {\rm d}{\bf x} \, \delta(E- H({\bf x}))$ the
volume of the constant--energy surface in 
phase space.
The brackets $\langle \dots \rangle_{\Delta E}$ denote a smooth 
(e.g., Gaussian) energy average over an energy window $\Delta E$ 
much larger than the mean level spacing but classically small,
i.e., $\langle d (E)\rangle_{\Delta E}^{-1} \ll \Delta E \ll E$. 
The mean density of states
determines the Heisenberg time,
\begin{equation}
\label{eq-heisenberg}
T_H \equiv 2 \pi \hbar \left\langle d(E) \right\rangle_{\Delta E} \, .
\end{equation}
As shown in Ref.~\cite{Prange97}, a  second average is required 
to obtain a self--averaging quantity for
the Fourier transform of the correlation function $R_{ab} (\epsilon)$. 
We thus introduce the form factor  as a time average of  
the Fourier transform of $R_{ab} (\epsilon)$  
over a time window
$\Delta T$, with $\Delta T \ll T_H$ (for instance, $\Delta T = 2\pi \hbar / \Delta E)$.
Denoting by $h(\epsilon)$ the Fourier transform of the
weight function in the time average, we define
\begin{equation} \label{eq-form_factor}
K_{ab} (T ) 
 \equiv 
  \langle d (E) \rangle_{\Delta E} \int\limits_{-\infty}^\infty {\rm d} \epsilon 
   \, e^{- i \epsilon T/\hbar} \, 
     h(\epsilon) \,
     R_{ab} (\epsilon)\,.
\end{equation}
This generalized form factor which has been introduced in 
Ref.~\cite{Eckhardt95}
will be the central quantity of this paper.
For definiteness, we consider
a uniform average over a time window 
$[T - \Delta T/2, T + \Delta T/2]$ implying
$h(\epsilon) =  (\epsilon \Delta T/2 \hbar)^{-1} \sin ( \epsilon \Delta T/2 \hbar)$. 

Setting $\hat{a} = \hat{b}= \hat{1}$
in Eqs.~(\ref{eq-correl_function}) and (\ref{eq-form_factor}),
one recovers the well--known {\em spectral} form factor $K(T) \equiv K_{11} (T)$. 
The correlation function $R(\epsilon)$ and its Fourier transform
(\ref{eq-form_factor}) have been calculated based on random--matrix 
assumptions.
For systems with time--reversal symmetry, 
the relevant random--matrix ensemble  is the 
Gaussian orthogonal ensemble (GOE) and yields
the spectral form factor
\begin{eqnarray} \label{eq-K_GOE}
\nonumber
K^{\text{}}(\tau)  
& =  &  2 \tau - \tau \ln ( 1 + 2 \tau ) \;\;\;, \;\;\; 0 < \tau < 1
\\  
& = &  2 \tau - 2 \tau^2 + {\mathcal{O}} (\tau^3)  \;,
\end{eqnarray}
independent of the dimensionality $f$ of the system.
Here, $K(\tau)$ is expressed in terms of the rescaled
time $\tau = T/T_H$.

As follows from Snirelman's theorem~\cite{Snirelman74}, 
the corresponding generalized form factor reads, to leading
order in $\hbar$, 
\begin{equation}
\label{leading_K_ab}
K_{ab}(\tau) \simeq \overline{a({\bf x} )} \; \overline{b({\bf x} )} \, K(\tau)
\end{equation}
(see section~\ref{sec:matrix-element_fluct} below).
Here, the average $\overline{a({\bf x})}$ of
the Weyl symbol $a({\bf x})$ of the quantum
observable $\hat{a}$
is taken with respect to the Liouville measure,
\begin{equation}
\label{def:ps_average}
 \overline{a({\bf x})} \equiv \frac{1}{\Omega}
 \int {\rm d} {\bf x} \, \delta(E- H({\bf x})) \, a({\bf x}) \, ,
\end{equation}
see also (\ref{eq-Weylsymbol}).
The phase--space coordinates are denoted by ${\bf x} = ({\bf q},{\bf p})$.
In many interesting situations $\overline{a({\bf x})} = \overline{b({\bf x})} = 0$
which can always be obtained by shifting 
$a({\bf x}) \to a({\bf x}) - \overline{a({\bf x})}$.
This implies, according to Eq.~(\ref{leading_K_ab}),
a vanishing $K_{ab}(\tau)$ for $\hbar \to 0$.
In this case, our semiclassical methods will enable us to go beyond the result
(\ref{leading_K_ab}).
We will show that the correction terms to Eq. (\ref{leading_K_ab})
are of order $1/T_H \sim \hbar^{f-1}$ and  given by  
\begin{equation}
\label{result_K_ab}
  K_{ab} \left( \tau \right) \approx \frac{1}{\tau T_H} \,
  \left[ 2 \tau - 2 \tau^2 + {\cal O}(\tau^3) \right] \,
  \int\limits_0^{\infty} \diff t \, C^S_{ab}(t)\;.
\end{equation}
Here the classical time--correlation function $C^S_{ab}(t)$ is
defined as
\begin{equation}
\label{def:correl_ab}
 C^S_{ab}(t) \equiv \overline{a({\bf x}) b^S({\bf x}_t)} 
 \quad \text{with} \quad b^S({\bf x}) = \frac{b({\bf x}) + b({\cal T}{\bf x})}{2}\;,
\end{equation}
where $\overline{a({\bf x})} = \overline{b({\bf x})}=0$ is implicitly
used, $\T : ({\bf q}, {\bf p}) \mapsto ({\bf q}, - {\bf p})$
is the time--reversal map, 
and ${\bf x}_t$ is the solution of the classical equations of motion with
initial condition ${\bf x}_0={\bf x}$. The symmetrized form $b^S({\bf x})$ of 
$b({\bf x})$ enters as the dynamics is assumed to be
invariant with respect to time reversal. 

\subsection{Semiclassical limit}
\label{subsec-semicl_K}

In the semiclassical approach,
the Weyl symbol of the operator $\hat{a}$,
\begin{equation} \label{eq-Weylsymbol}
 a({\bf q},{\bf p} ) \equiv 
 \int \D {\bf q} ' \, e^{\I\,  {\bf p} \cdot {\bf q}' /\hbar} \,
 \Bigl\langle {\bf q} - \frac{{\bf q} '}{2} \Bigl| \, \hat{a} \, \Bigr|  
  {\bf q} + \frac{ {\bf q}'}{2} \Bigr\rangle
\end{equation}
plays an important role (see, e.g., Ref.~\cite{Balazs84}).
It  is a function of the phase--space coordinates ${\bf x}=({\bf q},{\bf p})$ and
tends in the limit $\hbar \to 0$ to the corresponding classical observable.   
In the following we assume that $a({\bf x})$ is a smooth function of
${\bf x}$. The semiclassical evaluation
of Eq.~(\ref{gen_density}) for classically chaotic
quantum systems yields the generalized Gutzwiller trace 
formula~\cite{Eckhardt92,Combescure99}
\begin{equation}
\label{total_sc_density}
d_a(E) = \left\langle d_a(E) \right\rangle_{\Delta E} + d_a^{\rm osc}(E) \;,
\end{equation}
with
\begin{equation}
\label{sc_density_mean}
  \left\langle d_a(E) \right\rangle_{\Delta E} = \frac{\Omega}{(2 \pi \hbar)^{f}}
  \; \overline{a({\bf x})} 
\end{equation}
and
\begin{equation}
\label{sc_density_osc}
  d^{\rm osc}_a(E) = \frac{1}{\pi \hbar} 
 {\rm Re} \left\{
 \sum_\gamma {\cal A}_\gamma \exp\left({\rm i} \frac{S_\gamma}{\hbar} \right)
 \right\}.
\end{equation}
The mean weighted density of states,
$\left\langle d_a(E) \right\rangle_{\Delta E}$,
depends on the dimensionality $f$ of the system and
is determined by
the average (\ref{def:ps_average}) of $a({\bf x})$ over the constant--energy surface.
The oscillatory contribution, Eq.~(\ref{sc_density_osc}), is given by
a sum over classical periodic orbits labeled by $\gamma$.
The  weights ${\cal A}_\gamma$ are related to
the amplitudes~\cite{Gutzwiller90}
\begin{equation} \label{eq-w} 
w_\gamma = \frac{(T_\gamma / r_{\gamma})
 \exp(-{\rm i} \pi \mu_\gamma /2)}{\sqrt{|\det(M_\gamma - 1)|}}
\end{equation}
 via the relation
${\cal A}_\gamma = w_\gamma A_\gamma$, where $T_\gamma$
is the period of the orbit $\gamma$,
$r_\gamma$ its repetition number, 
$\mu_\gamma$ its Maslov index, $M_\gamma$ its stability matrix, and
\begin{equation}
\label{A_gamma}
 A_\gamma = A({\bf x}_0^\gamma, T_\gamma) 
 \equiv \frac{1}{T_\gamma} \int\limits_0^{T_\gamma} \, {\rm d}t \,
 a({\bf x}_t^\gamma)\;.
\end{equation}
Here, ${\bf x}_t^\gamma$ is the phase--space point on the periodic orbit $\gamma$ 
obtained by solving the classical equations of motion with
the initial condition ${\bf x}_0$,
so that ${\bf x}^\gamma_{t+T_\gamma} = {\bf x}^\gamma_t$.
The applicability of the semiclassical 
expression (\ref{sc_density_osc}) to chaotic systems with more than
two degrees of freedom has been extensively studied in
Ref.~\cite{Primack00}. 

Only the oscillating parts of $d_a(E)$ and $d_b(E)$ contribute to the correlation function
(\ref{eq-correl_function}). 
Substituting Eq.~(\ref{sc_density_osc}) into Eq.~(\ref{eq-form_factor}),
one obtains
\begin{widetext}
\begin{equation}
\label{gen_K}
 K_{ab}(T) = \frac{1}{T_H} \left\langle\sum\limits_{\gamma, \bar{\gamma}}
 {\cal A}_\gamma {\cal B}_{\bar{\gamma}}^* \;  
 \exp \left({\rm i} \frac{S_\gamma - S_{\bar{\gamma}}}{\hbar} \right) \;
 \delta_{\Delta T}\left(T - \frac{T_\gamma + T_{\bar{\gamma}}}{2}\right)
 \right\rangle_{\Delta E}
\end{equation}
\end{widetext}
by generalizing the corresponding steps of the semiclassical
derivation of the
{\em spectral} form factor~\cite{Haake}.
In Eq.~(\ref{gen_K}), the delta function with a finite width, 
$\delta_{\Delta T} (T')$, originates from our choice of time averaging
in Eq.~(\ref{eq-form_factor}). It is equal to 
$\Delta T^{-1}$ if $-\Delta T / 2 \leq T' \leq \Delta T / 2$
and zero otherwise. The semiclassical formula (\ref{gen_K})
of $K_{ab}(T)$ is the starting point of a further semiclassical
evaluation. It contains a double sum over terms
which strongly fluctuate with energy and poses the
challenge to approximately compute its energy average.

An earlier approach to this problem is presented in Ref.~\cite{Bogomolny96}
where the correlation function (\ref{eq-correl_function}) is considered directly
instead of the form factor. It was shown that the off--diagonal contributions
can be related to the diagonal terms yielding the leading order oscillatory
term of the corresponding RMT result of $R(\epsilon)$ for large
$\epsilon$. Furthermore, it was pointed out that in the case of a 
weighted density of states 
these contributions vanish to leading semiclassical order if the 
microcanonical average of the corresponding observable is zero.
In this work we proceed in a different way and
restrict our considerations to the
form factor in the limit of small $\tau=T/T_H \ll 1$, see
Subsection \ref{the_leading_term} for
our corresponding results concerning the weighted density of states.

In the following section we first discuss
the case of the spectral form factor and then generalize
our approach to include matrix element fluctuations in
Section~\ref{sec:matrix-element_fluct}.

\section{Spectral form factor for $f$-dimensional systems}
\label{sec:higher_dimensional}

\subsection{Semiclassical evaluation within the diagonal approximation} 
\label{sec-diag_approx}

A semiclassical expression for the spectral form factor
$K(\tau) = K_{11}(\tau)$ is given by Eq.~(\ref{gen_K})
with
${\cal A}_\gamma = w_\gamma$, ${\cal B}_{\bar{\gamma}} = w_{\bar{\gamma}}$
and the rescaled time $\tau = T/T_H$.
To leading order in $\hbar$ and $\tau$, the double sum over periodic
orbits can be evaluated
by means of the so--called diagonal approximation~\cite{Berry85}.
It is guided by the fact that the contributions from pairs $(\gamma,\overline{\gamma})$
of orbits with 
action differences larger than $\hbar$ strongly fluctuate in
energy and are therefore suppressed upon energy averaging.
Hence the main contribution
stems from the pairs of orbits with equal actions 
$S_\gamma = S_{\bar{\gamma}}$. 
If the system has no other than time--reversal symmetry then
these pairs are obtained (up to accidental action
degeneracies) by pairing 
each orbit $\gamma$ with itself
or with its time--reversed version $\gamma^i$. 
To calculate the corresponding contribution
to the semiclassical form factor (\ref{gen_K}), 
one has to perform a weighted periodic--orbit average of the type
\begin{equation}
\label{po_average}
 \left \langle \dots \right \rangle_{{\rm po},T} \equiv
 \frac{1}{T} \sum\limits_\gamma \dots |w_\gamma|^2 \delta_{\Delta T}
 (T-T_\gamma).
\end{equation}
This is achieved by means of the following sum rule for periodic
orbits in chaotic systems~\cite{Parry90}:
\begin{equation}
\label{sum_rule}
 \left\langle \frac{1}{T_\gamma} \int\limits_0^{T_\gamma} {\rm d}t \, f({\bf x}_t^\gamma)
 \right\rangle_{{\rm po},T} \simeq \frac{1}{T}  \int\limits_0^{T} {\rm d}t \, f({\bf x}_t)
 \simeq \overline{f({\bf x})}
\end{equation}
with $T \to \infty$.
On the left--hand side the arbitrary continuous function
$f({\bf x})$ is integrated along 
the periodic orbits $\gamma$ with periods $T_\gamma$ lying in 
the interval $[T - \Delta T / 2, T+ \Delta T / 2]$.
The integral on the right--hand side of Eq.~(\ref{sum_rule}) 
is taken along a nonperiodic ``ergodic'' trajectory which
uniformly and densely fills the constant--energy surface.
For ergodic systems this time average
is equal to the phase--space average $\overline{f({\bf x})}$
in the large $T$ limit.  In the special case $f({\bf x})=1$, 
Eq.~(\ref{sum_rule}) is known as the Hannay-Ozorio de Almeida sum 
rule~\cite{HODA84}.

For a fixed rescaled time $\tau=T/T_H$, the
periods $T_\gamma$ of the orbits entering Eq.~(\ref{gen_K})
diverge as $T_H = {\mathcal{O}} ( \hbar^{1-f} )$ for $\hbar \to 0$.
This justifies the use of the sum rule (\ref{sum_rule}) for the evaluation
of the form factor. Since $w_\gamma = w_{{\gamma}^i}$,
see Eq.~(\ref{eq-w}),
the contribution of the pairs $(\gamma,\gamma)$ and $(\gamma, \gamma^{i})$
to the spectral form factor, neglecting all other orbit 
pairs, is given by~\cite{Berry85}
\begin{equation}
\label{K_diag}
        K^{(1)} \left(\tau \right) = 2  \tau  
        \left\langle\left\langle 1 \right\rangle_{{\rm po},T}
        \right\rangle_{\Delta E} \simeq 2 \tau \; .
\end{equation}
The factor of two is due to the time--reversal symmetry.
It is worthwhile to note that $K^{(1)}(\tau)$ agrees with the leading
term of the RMT result (\ref{eq-K_GOE}) for the
GOE case; correspondingly the GUE form factor $K^{(1)}(\tau)=\tau$
is reproduced for systems without time--reversal symmetry.
Hence the diagonal approximation 
explains the universality of the form factor for $\tau \ll 1$ in the
semiclassical limit. 

\subsection{Origin of the off--diagonal corrections}
\label{sec-off_diag}

\begin{figure}
\centerline{
\epsfxsize=0.4\textwidth
\epsfbox{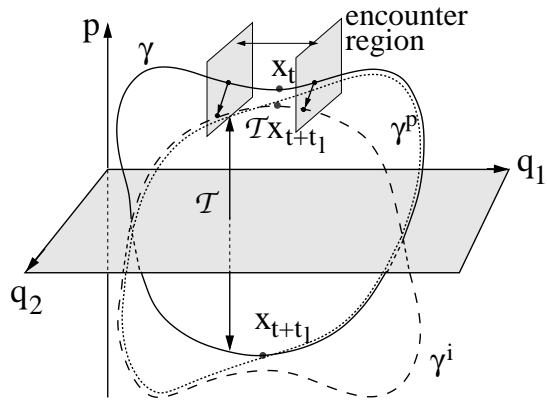}}
\caption{
\label{fig:encounter}
Representation of a periodic orbit $\gamma$ (solid line)
in phase space with a close encounter together
with its time--reverse version $\gamma^i$ (dashed line) and its partner orbit $\gamma^p$
(dotted line). The orbit $\gamma$  is characterized by two stretches
which are almost time reverse of one another. One of these stretches
is situated between the two 
Poincar{\'e} surface of sections (PSS) perpendicular to the orbit shown by the grey squares. The time--reverse map $\T$ is
the reflection with respect to the plane $p=0$. The picture should be thought of as
a projection  of the whole 
$2f$-dimensional phase space on a subspace formed by
one  momentum and two position coordinates. 
}
\end{figure}
In order to go beyond the diagonal approximation and to explain the agreement of the 
semiclassical spectral form factor with 
the RMT result (\ref{eq-K_GOE})  at higher orders in $\tau$, one has to evaluate further
terms in the double sum over periodic orbits (\ref{gen_K}). Only
pairs of periodic orbits which involve
a small action difference of the order of $\hbar$ interfere constructively 
and are not suppressed by the energy average.
In a series of papers~\cite{Spehner03,Turek03,Mueller03,Sieber01,Sieber02} starting from
the work by Sieber and Richter~\cite{Sieber01,Sieber02}, specific 
periodic--orbit 
correlations in systems with time--reversal symmetry have been investigated
in order to compute the 
leading off--diagonal corrections to the semiclassical spectral form factor.
It has been shown that,  for hyperbolic dynamics invariant under time reversal, 
there exists a continuous family of
pairs $(\gamma, \gamma^p)$ of periodic orbits with arbitrarily 
small action differences. These orbit pairs give rise to a contribution of
$K^{(2)} (\tau) = - 2 \tau^2$ to the spectral form factor. Hence it 
coincides with the next-to-leading-order term in the RMT result (\ref{eq-K_GOE}). 
The idea of the approach runs as follows.
A periodic orbit $\gamma$ is represented by a closed 
curve in phase space. Let us assume that this curve has
two stretches which are almost the time--reverse of one another (i.e., they are almost 
identical after applying the time reverse map 
$\T : ({\bf q},{\bf p}) \mapsto ({\bf q}, - {\bf p})$ on one of them). In the sequel, we
refer to such two almost time--reverse stretches of the orbit $\gamma$ as a 
``close encounter'' or, more shortly, an ``encounter''.
The two pieces of $\gamma$
separated by the encounter are called the ``left part'' and the ``right part''.   
One can associate to $\gamma$ a 
partner orbit $\gamma^p$ by inverting time on, say, the left part, leaving the 
right part almost unchanged. Hence  $\gamma^p$ follows closely  the time reverse 
$\gamma^i$ of $\gamma$ on its left part while it  follows closely $\gamma$   on its right part,
as shown in Fig.~\ref{fig:encounter}.
Such a partner orbit $\gamma^p$ has
almost the same action as $\gamma$. More precisely, the more symmetric
the two orbit stretches of $\gamma$ are  with respect
to time reversal the closer is $\gamma^p$ to either
$\gamma$ or $\gamma^i$ and the
smaller the action difference $S_{\gamma^p}-S_\gamma$. 
Because the periods of the orbits involved
in the sum (\ref{gen_K}) are on the scale
of the Heisenberg time $T_H$, Eq.~(\ref{eq-heisenberg}), one expects a large number
of encounters on a given orbit $\gamma$ in this sum. 
Thus a large number of partner orbits $\gamma^p$ with small action 
differences $S_{\gamma^p}-S_\gamma$ can be associated to any periodic
orbit $\gamma$.
Both, $\gamma$ and its associated partner orbit $\gamma^p$
share the property to have 
two almost time--reverse stretches, which are approximately the same for 
both orbits. The partner orbit of $\gamma^p$ coincides with the original orbit $\gamma$.

\begin{figure}
\centerline{
\epsfxsize=0.4\textwidth
\epsfbox{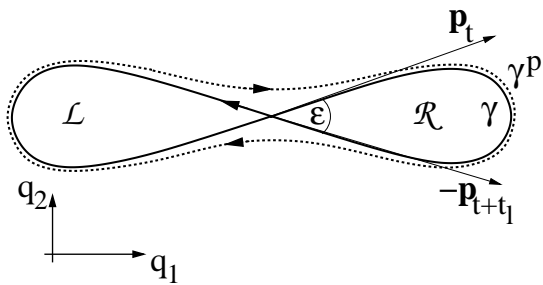}}
\caption{
\label{fig:SR_loops}
Configuration space representation of a 
periodic orbit $\gamma$ with a close encounter (solid line) together
with its partner orbits $\gamma^p$
(dotted line) for a system with two degrees of freedom. 
}
\end{figure}

All previous
works~\cite{Sieber01, Sieber02, Spehner03, Turek03, Mueller03, Mueller04} dealing with the 
contribution of the pairs  $(\gamma,\gamma^p)$ of partner orbits to the semiclassical form factor  
have been  restricted to 
systems with two degrees of freedom. Then either  
$\gamma$ or $\gamma^p$ has one (or possibly several)
self--intersection(s) in configuration space,
which corresponds to the encounter in phase space. 
The right and left parts of the orbit correspond to the two loops 
formed by this intersection in configuration space,
see Fig.~\ref{fig:SR_loops}. The right loop is traversed in the
same direction while the left loop is traversed with different
orientation, hence requiring time--reversal symmetry.
In order that the two stretches of the orbit
near the self--intersection
be almost symmetric with respect to time reversal, 
the two corresponding velocities must be almost antiparallel. The intersection 
is then characterized 
by a small crossing angle $\varepsilon $.  
The orbits $\gamma$ and $\gamma^p$ are
distinguished by the fact that one has one more self--intersection 
than the other~\cite{Mueller03}.
This is in contrast to the phase--space approach, 
in which the two partner orbits are treated on equal footing.
In systems with more than two degrees of freedom, the phase--space 
approach~\cite{Spehner03, Turek03} is  more appropriate, because
for $f>2$
the relevant orbits generally do not have self--intersections
in the $f$-dimensional configuration space.

In the following subsections, we will study the spectral form factor of 
quantum mechanical systems whose
classical counterparts are Hamiltonian systems with
$f \geq 2$ degrees of freedom. Furthermore, we
consider systems with time--reversal
symmetry, since only in this case the 
orbit pairs $(\gamma,\gamma^p)$ exist. 
We show that, if the underlying classical dynamics is  ergodic and hyperbolic,
these orbit pairs yield the contribution 
$K^{(2)} (\tau) = - 2 \tau^2$ to the semiclassical spectral form 
factor, independent of the number of degrees of freedom.
Remarkably, the different time scales given
by the set of Lyapunov exponents 
$\{ \lambda_i \}$ do not show up in the final result
which coincides with the universal second--order term of the random--matrix theory
prediction (\ref{eq-K_GOE}). Our technique strongly relies on the
equivalence between the two approaches previously developed
in Ref.~\cite{Turek03}
and Refs.~\cite{Spehner03,Heusler03} to count the number of partner orbits.
Therefore we present a proof of this equivalence
which clarifies the underlying dynamical mechanisms
related to the partner orbit statistics. Our semiclassical
evaluation of the spectral form factor
$K^{(2)}(\tau)$ will serve as a basis for the calculation
of the generalized form factor $K_{ab}^{(2)}(\tau)$
in Section~\ref{sec:matrix-element_fluct}.

\subsection{Hyperbolic Hamiltonian systems}
\label{subsec:hyperbolic_hamiltonian}

Before evaluating the spectral form factor we introduce the notations
by very briefly summarizing the necessary concepts
for dynamical systems~\cite{Gaspard98}. The classical dynamics
of the system is assumed to be ergodic and hyperbolic. It maps any phase--space
point ${\bf x}_0 = {\bf x}$ onto the point ${\bf x}_t$
after time $t$. Hyperbolicity 
means that all Lyapunov exponents are nonzero except
the one corresponding to the direction along the
flow~\cite{Gaspard98}.  For a given classical trajectory, the dynamics in its
vicinity can be linearized using the
stability matrix $M(t, {\bf x})$. The vector
$\delta \vec{y}_0 ( {\bf x} ) \equiv (\delta {\bf q}^\perp_0, \delta {\bf p}^\perp_0)$
describing a small displacement from ${\bf x}$
perpendicular to the trajectory\footnote{
We will specify displacement vectors in the
$2(f-1)$--dimensional PSS
by using an arrow, e.g., $\delta \vec{y}$, while vectors in
the $2f$-dimensional phase space are written
in bold face, e.g., ${\bf x}$.
} 
within the constant--energy surface
is given at a later time $t$ by
\begin{equation}
\label{lin_motion}
  \delta \vec{y}_t({\bf x}) \approx 
  M(t,{\bf x}) \; \delta \vec{y}_0({\bf x}).
\end{equation}
This linear approximation is valid as long as 
$\delta \vec{y}_t({\bf x})$ remains sufficiently small. 
The set of all possible vectors $\delta \vec{y}_{0}({\bf x})$ defines a $(2f-2)$-dimensional
Poincar{\'e}
surface of section (PSS) at point ${\bf x}$ perpendicular
to the trajectory in phase space. The matrix $M(t,{\bf x})$ is a linear map from the
PSS at ${\bf x}$
to the PSS at ${\bf x}_t$.
This map is 
symplectic, i.e., it satisfies $M^T\; \Sigma \; M = \Sigma$, with
\begin{equation}
  \Sigma = \left(
  \begin{array}{c c} {\bf 0} & {\bf 1} \\-{\bf 1} & {\bf 0} \end{array}
  \right) \; ,
\end{equation}
where ${\bf 0}$ and ${\bf 1}$ refer to the
$(f-1) \times (f-1)$ null and identity  matrices.
Therefore the symplectic product is conserved by the dynamics, i.e., 
$\delta \vec {y}_t^{\, T} \; \Sigma \; \delta {\vec{y}_t}^{\,\prime} \approx 
\delta \vec{y}_0^{\, T} \; \Sigma \; \delta {\vec{y}_0}^{\,\prime}$ for any
two small displacements $\delta \vec{y}_{0}$ and $\delta {\vec{y}_{0}}^{\,\prime}$, provided
$\delta \vec{y}_t$ and $\delta {\vec{y}_t}^{\, \prime}$ remain sufficiently small.
 
The linear stable and unstable directions in the
PSS at ${\bf x}$ are denoted
by $\vec{e}^{\,s}_i({\bf x})$ and $\vec{e}^{\, u}_i({\bf x})$.
They define vector fields which can be found by means of a homological
decomposition~\cite{Gaspard98} of the stability matrix $M(t, {\bf x})$.
A stretching factor $\Lambda_i(t,{\bf x})$ is associated to each direction. It is defined by
\begin{equation} \label{eq-def_Lambda}
M(t,{\bf x}) \vec{e}^{\, u,s}_i({\bf x}) = \Lambda_i(t,{\bf x})^{\pm 1}
 \; \vec{e}^{\, u,s}_i({\bf x}_t) \;,
\end{equation}
where the signs $+$ and $-$ correspond to the
superscripts $u$ and $s$, respectively. 
It is worth noting that Eq.~(\ref{eq-def_Lambda}) is
not an eigenvalue equation for the matrix $M(t,{\bf x})$, since
the vectors $\vec{e}^{\, s,u}$ are evaluated at different positions in
phase space~\footnote{
Eq.~(\ref{eq-def_Lambda}) can be viewed as an eigenvalue equation 
only for points ${\bf x}= {\bf x}^\gamma$ belonging to periodic orbits $\gamma$ and
times $t$ equal to (a multiple of) the period $T_\gamma$. Then
$M(T_\gamma,{\bf x}^\gamma)=M_\gamma$ is the stability matrix of $\gamma$ at
${\bf x}^\gamma$ and
$| \Lambda_i ( T_\gamma, {\bf x}^\gamma ) | = \exp ( \lambda_i^\gamma \,T_\gamma )$
gives the Lyapunov exponents $\lambda_i^\gamma$ of the orbit.
}.
In the long--time limit, the stretching factor 
is related to the Lyapunov exponent $\lambda_i({\bf x})$ at point 
${\bf x}$ via the relation
$\ln | \Lambda_i (t, {\bf x} ) | \approx  \lambda_i({\bf x}) t $. On shorter time
scales one has to solve the equations of motion 
\begin{equation}
\label{eom_lambda}
        \frac{\diff \Lambda_i(t, {\bf x})}{\diff t} = \chi_i({\bf x}_t ) \,
        \Lambda_i(t, {\bf x})\;,
\end{equation}
where $\chi_i({\bf x})$ is the local growth rate. In the following,
we will assume that the local growth rates are continuously
varying functions in phase space. In general, $\chi_i({\bf x})$
can take negative values in some region of phase space~\cite{Gaspard98}.
However, by ergodicity, its average $\overline{\chi_i ({\bf x})}$
over the constant--energy surface is positive since it is equal to the 
$i$th positive Lyapunov exponent $\lambda_i$
at almost all points ${\bf x}$ (i.e., on a set of points ${\bf x}$ of measure one). 
The Lyapunov exponents of periodic orbits 
(being of measure zero in phase space)
are in general different from the $\lambda_i$'s. They 
are given by
\begin{equation} \label{eq-Lyapunov_exp}
\lambda_i^{\gamma} \equiv \lambda_i({\bf x}_0^\gamma) 
= \frac{ \ln | \Lambda_i (T_\gamma,{\bf x}_0^\gamma) |}{T_\gamma}
= \frac{1}{T_\gamma}
\int_0^{T_\gamma} \D t \,\chi_i ( {\bf x}_t^\gamma)\;.
\end{equation}

In hyperbolic systems, the set of vectors 
$\{\vec{e}^{\, s}_i({\bf x}), \vec{e}^{\, u}_i({\bf x})\}$ 
spans the whole
PSS at ${\bf x}$. Hence
each displacement vector $\delta \vec{y}({\bf x})$ can be decomposed into
its stable and unstable components,
\begin{equation}
\label{y_components}
  \delta \vec{y}({\bf x})  \equiv \delta \vec{y}_s ({\bf x}) + \delta \vec{y}_u ({\bf x})=
  \sum\limits_{i=1}^{f-1} \Bigl( 
   s_i ({\bf x})\, \vec{e}^{\, s}_{\, i}({\bf x}) +
   u_i({\bf x}) \, \vec{e}^{\, u}_{\, i}({\bf x}) \Bigr)  \,.
\end{equation}
Therefore $\delta \vec{y}({\bf x})$ is determined by the set of 
stable coordinates $\{s_i\}$ and unstable coordinates $\{u_i\}$.
Provided that all these coordinates $u_i, s_i$ are small enough, 
the linear approximation (\ref{lin_motion})
can be applied for sufficiently long times $t$, say, up to some time
$\Delta t_u \gg \lambda_i^{-1} ({\bf x})$. By (\ref{eq-def_Lambda}) and (\ref{y_components}), 
the $i$th unstable component
at time $t$ is then equal to its value $u_i$ at time $t=0$ multiplied 
by $\Lambda_i (t,{\bf x})$.
This leads to an exponential
growth of this unstable component as $|u_i | \exp [\lambda_i({\bf x})  t]$
during the time $\lambda_i^{-1}({\bf x}) \ll t \leq \Delta t_u$. Similar arguments
hold for the stable components $s_i$ when going backwards in time.
This implies an exponential decrease of $s_i$ so that
the product $u_i \,s_i$ remains constant.
For times $t \leq \Delta t_u$, it follows from Eq.~(\ref{eom_lambda}) that
\begin{equation} \label{chi}
\frac{\D u_i}{\D t}=\chi_i({\bf x}_t) \, u_i \; 
\end{equation}
and similarly for $s_i$ with $\chi_i$ replaced by $-\chi_i$.

The dynamics uniquely specifies the directions
of the $\vec{e}^{\, s,u}_i({\bf x})$. Due to the 
symplectic nature of the stability matrix $M(t,{\bf x})$,
they have to fulfill the ``orthogonality relations''
\begin{subequations}
\label{orthogonality+units}
\begin{eqnarray}
\label{orthogonality}
  \vec{e}^{\, u}_i({\bf x})^T \, \Sigma \, 
  \vec{e}^{\, u}_j({\bf x}) & = &
  \vec{e}^{\, s}_i({\bf x})^T \, \Sigma \,
  \vec{e}^{\, s}_j({\bf x}) \nonumber \\
  & = &
  \vec{e}^{\, u}_i({\bf x})^T \, \Sigma \, \vec{e}^{\, s}_j({\bf x}) = 0
\end{eqnarray}
for $i \not= j$.
However, the  norms of 
$\vec{e}_i^{\, s,u}({\bf x})$  can be chosen arbitrarily. 
In the sequel, we choose these norms in a way
that their symplectic product gives a
classical action $S_{\text{cl}}$ of the system
under consideration, e.g., the action
corresponding to the shortest periodic orbit,
\begin{equation}
\label{units}
  \vec{e}^{\, u}_i({\bf x})^T \, \Sigma \, 
  \vec{e}^{\, s}_j({\bf x})
  = S_{\text{cl}} \delta_{ij}.
\end{equation}
\end{subequations}
The symmetry of the dynamics with respect to the time--reversal 
operation $\T$ implies
$M( t, \T {\bf x}_t )  = \T M( t, {\bf x} )^{-1} \,\T$, where,
in the right--hand side, the symbol $\T$ refers to the  
restriction of the time--reversal map to the
PSS at ${\bf x}$ or at $\T {\bf x}_{t}$. It follows that the vectors 
$\vec{e}^{\, s,u}_i({\bf x})$ can be chosen in such a way that, in addition to Eq.~(\ref{orthogonality+units}),
they satisfy
\begin{subequations}
\label{eq-time_reverse_e^u,s}
\begin{equation}
  \vec{e}^{\, u}_i( \T {\bf x})  =  \T \vec{e}^{\, s}_i( {\bf x}) 
 \; , \quad 
 \vec{e}^{\, s}_i( \T {\bf x}) =  \T \vec{e}^{\, u}_i( {\bf x}) 
\end{equation}
and
\begin{equation}
\Lambda_i(t, \T {\bf x}_t ) = \Lambda_i ( t, {\bf x})\;. 
\end{equation}
\end{subequations}
The relations (\ref{orthogonality+units}) imply
that the Jacobian matrix 
$J({\bf x} ) = 
\partial(  \delta {\bf q}^\perp , \delta {\bf p}^\perp )/\partial ( s_i, u_i )$
of the transformation from the position/momentum coordinates to the stable/unstable
coordinates in the PSS at ${\bf x}$ 
is symplectic up to a factor $-S_{\text{cl}}$,
i.e., $J({\bf x} )^T \Sigma J({\bf x} ) = -S_{\text{cl}} \Sigma$.
Hence the Jacobian determinant of this transformation equals $(-S_{\text{\text{cl}}})^{f-1}$.

\subsection{Encounter region}
\label{subsec:encounter}

\begin{figure}
\centerline{
\epsfxsize=0.4\textwidth
\epsfbox{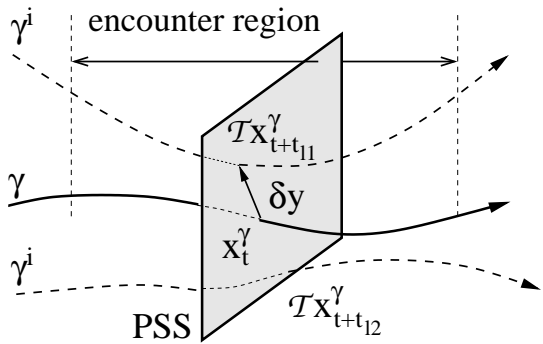}}
\caption{
\label{fig:encounter_region}
Schematic drawing of the encounter region in phase space.
The Poincar{\'e} surface of section (PSS) is
a $(2f-2)$-dimensional surface defined
at ${\bf x}^\gamma_t$ in the $2f$-dimensional phase
space by the perpendicular coordinates
$(\delta {\bf q}^\perp, \delta {\bf p}^\perp)$.
The original orbit $\gamma$ is represented by the solid line.
Also drawn are two segments of the time--reversed 
orbit $\gamma^i$ (dashed lines) which yield the
two closest intersection points. The corresponding
``loop times'' $t_{\rm l}$ are denoted by $t_{\rm l1}$ and
$t_{\rm l2}$. The displacement vector 
$\delta \vec{y} = \delta \vec{y}({\bf x}_t^\gamma, t_{\rm l})$ points from the
original orbit to the intersection points.
}
\end{figure}

As it has been described in Subsection~\ref{sec-off_diag} the  pairs of periodic
orbits $(\gamma,\gamma^p)$ which interfere constructively in the double sum  
(\ref{gen_K}) are related to close encounters of $\gamma$.
Each such encounter involves two orbit 
stretches of $\gamma$ which are approximately 
time reversed with respect to each other.
The purpose of this subsection is to give a 
more precise and quantitative definition of the notion
of an encounter in the $2 f \geq 4$ dimensional phase space.

Let us assume that the periodic orbit $\gamma$ comes close to its time--reversed version
$\gamma^i$ at a point ${\bf x}^\gamma_t$ in phase space
so that ${\bf x}^\gamma_t \simeq {\cal T}{\bf x}_{t+t_{\rm l}}^\gamma$. 
In the following, we choose the time $t_{\rm l}$ such
that ${\cal T}{\bf x}_{t+t_{\rm l}}^\gamma$ lies in the
PSS perpendicular to the orbit
at ${\bf x}^\gamma_t$.
Thus the small displacement vector between $\gamma$ and
$\gamma^i$ lies in this PSS 
and can be decomposed in terms of the stable and unstable coordinates 
(see Subsection~\ref{subsec:hyperbolic_hamiltonian}),
\begin{eqnarray} 
\label{delta_y}
 \delta \vec{y}({\bf x}_t^\gamma , t_{\rm l}) 
 & = & {\cal T}{\bf x}_{t+t_{\rm l}}^\gamma - {\bf x}_t^\gamma \\
  & = & \sum\limits_{i=1}^{f-1} \Bigl( s_i({\bf x}_t^\gamma , t_{\rm l})
   \vec{e}_i^{\, s}({\bf x}_t^\gamma ) + 
 u_i({\bf x}_t^\gamma , t_{\rm l}) \vec{e}_i^{\, u}({\bf x}_t^\gamma ) \Bigr) \;.
 \nonumber
\end{eqnarray}
If one moves from ${\bf x}_t^\gamma$ to ${\bf x}_{t+\Delta t}^\gamma$
along the orbit $\gamma$, 
this displacement vector evolves according to the
equations of motion and becomes  
$\delta \vec{y}_{\Delta t}({\bf x}_t^\gamma , t_{\rm l}) =
 \delta \vec{y}({\bf x}^\gamma_{t+\Delta t} , t_{\rm l} - 2 \Delta t)$.
The displacement $\delta \vec{y}_{\Delta t}({\bf x}_t^\gamma , t_{\rm l})$ 
remains small due to the deterministic nature of the
dynamics if the time $\Delta t$ is sufficiently short.
In other words, if the two orbits $\gamma$
and $\gamma^i$ are close to each other at some
point in phase space, it takes them a certain finite
time until they have significantly deviated from each other.

We define the ``encounter region'' as the set of all
points ${\bf x}_{t+\Delta t}^\gamma$  such that each stable and unstable
component 
of the displacement vector $\delta \vec{y}_{\Delta t}({\bf x}_t^\gamma,t_{\rm l})$
is smaller than a certain threshold $c \lesssim 1$. 
The value of $c$ is chosen in such a way that 
$\delta \vec{y}_{\Delta t}({\bf x}_t^\gamma , t_{\rm l})$
is given by the linearized equations of motion 
$\delta \vec{y}_{\Delta t}({\bf x}_t^\gamma , t_{\rm l}) 
 \approx M(\Delta t, {\bf x}_t^\gamma) \delta \vec{y}({\bf x}_t^\gamma, t_{\rm l})$
as long as ${\bf x}_{t+\Delta t}^\gamma$  stays within the encounter region, while
the linear approximation breaks down outside of it.
Therefore $c$ is a purely classical quantity which 
describes the breakdown of the linear approximation
applied to $\delta \vec{y}_{\Delta t}({\bf x}_t^\gamma, t_{\rm l})$.
As it will turn out, the precise value of $c$ is not essential
for the calculation of the form factor in the semiclassical limit.
This implies that a phase--space dependent $c({\bf x})$ does not alter
the final result for the form factor.
Strictly speaking, $c$ also depends on the size of the encounter region,
since 
the corrections to the linear approximation specified above should
increase with $\Delta t$. However, for smooth dynamics this time dependence
turns out to be  weak, i.e., logarithmic in $|u_i|,|s_i|$, and one can show that it
does not affect the result for the form factor~\cite{Spehner03}.

From the definition given above,
one concludes that the range of 
values of $\Delta t$ such that ${\bf x}_{t + \Delta t}^\gamma$ lies
within an encounter region is given by 
$- \Delta t_s \leq \Delta t \leq \Delta t_u$,
where $\Delta t_{s,u}$ is defined
as follows.
Let us denote the $i$th stable component
of the time--evolved vector 
$\delta \vec{y}_{\Delta t} ({\bf x}_t^\gamma , t_{\rm l})$ 
as $s_i({\Delta t}; {\bf x}_t^\gamma, t_{\rm l})$ and similarly for
the unstable components $u_i$. Then $\Delta t_u$ is such that
the displacement $\delta \vec{y}_{\Delta t_u}$
is just about to leave the hypercube 
${\cal C} = \left\{ (s_i, u_i): |s_i|,|u_i| \leq c \right\}$
meaning that its largest unstable component first reaches the
value $c$. A similar definition yields a time $\Delta t_s$ if going
backwards in time, so that
\begin{eqnarray}
\label{t_su}
  \max\limits_{i=1, \dots ,(f-1)}
  \left\{ | s_i({-\Delta t_s}; {\bf x}_t^\gamma , t_{\rm l}) | \right\} & = & c
  \quad \text{and}  \quad \nonumber \\
  \max\limits_{i=1, \dots , (f-1)}
  \left\{ | u_i({\Delta t_u}; {\bf x}_t^\gamma , t_{\rm l}) | \right\}  & = & c \, .
\end{eqnarray}
These implicit equations determine the times
$\Delta t_{s} = \Delta t_{s}(\{s_i\}; {\bf x}_t^\gamma )$ and 
$\Delta t_{u} = \Delta t_{u}(\{u_i\}; {\bf x}_t^\gamma )$
as functions of the components $\{s_i, u_i\}$ of
the vector $\delta \vec{y} ({\bf x}_t^\gamma , t_{\rm l})$ defined in (\ref{delta_y})
and of the point ${\bf x}_t^\gamma$ in phase space.
The time duration of the encounter region,
\begin{equation}
\label{t_enc}
 t_{\text{enc}}(\{s_i, u_i\}; {\bf x}_t^\gamma) 
 = \Delta t_u(\{s_i\}; {\bf x}_t^\gamma)
 + \Delta t_s(\{u_i\}; {\bf x}_t^\gamma)\;,
\end{equation}
thus depends on ${\bf x}_t^\gamma$ and on all the components 
$\{s_i, u_i \}$.
This time $t_{\text{enc}}$ is clearly invariant within
a given encounter region. 

The breakdown times $\Delta t_{s}$ and $\Delta t_{u}$
for the linearization can be estimated in the limit $|s_i|,|u_i| \ll 1$
by using Eq.~(\ref{t_su}) and the exponential growth of the unstable 
and stable components
in the forward and backward time directions, respectively.
With an error much smaller than 
$\Delta t_{u,s}$ themselves, they diverge like
$\Delta t_{s} \approx \lambda_j^{-1} \ln |s_j^{-1}|$
and $\Delta t_{u} \approx \lambda_k^{-1} \ln |u_k^{-1}|$
where $j$ and $k$ are the components for which the
maximal values are first reached.     
For $f > 2$ degrees
of freedom the presence of the maximum in Eq.~(\ref{t_su})
thus makes the functional dependence of $\Delta t_{u,s}$ and
$t_{\text{enc}}$ on 
$\{s_i , u_i \}$ rather complicated, in contrast to systems with 
two degrees of freedom.

\subsection{Partner orbit}
\label{subsec:partner_orbits}

Let us consider an orbit $\gamma$ having an encounter
at the phase--space location ${\bf x}^\gamma_{t}$ after time  $t_{\rm l}$
as described in the previous subsection. For now
we assume that the components $\{ s_i, u_i \}$
of the vector $\delta \vec{y} = \delta \vec{y} ({\bf x}_t^\gamma ,t_{\rm l})$
are small, i.e., $|s_i|, |u_i| \ll 1$.
As it will turn out in due
course this is the only relevant case for the form factor. 
We show, by analyzing the linearized 
equations of motion (\ref{lin_motion})
around $\gamma$ or $\gamma^i$,
that there exists another periodic orbit $\gamma^p$ which 
follows closely $\gamma$ between $t$ and $t+t_{\rm l}$ (part ${\cal R}$)
and follows closely $\gamma^i$ during the rest of the time (part ${\cal L}$), i.e.,
\begin{equation}
\label{partner_coords}
        {\bf x}_{t'}^{\gamma^p} \simeq \left\{
        \begin{array}{lll}
        {\bf x}_{t'}^\gamma & \;
        \text{for} \quad t \leq t' < t+t_{\rm l} & \; (\text{part }  {\cal R} ) \\ 
        {\cal T}{\bf x}_{2t+t_{\rm l}-t'}^\gamma & \; \text{for} \quad
        t+t_{\rm l} \leq t' < T_\gamma +t  & \; (\text{part } {\cal L} ). 
        \end{array} \right.
\end{equation}
%

\begin{figure}[t]
\centerline{
 \epsfig{figure={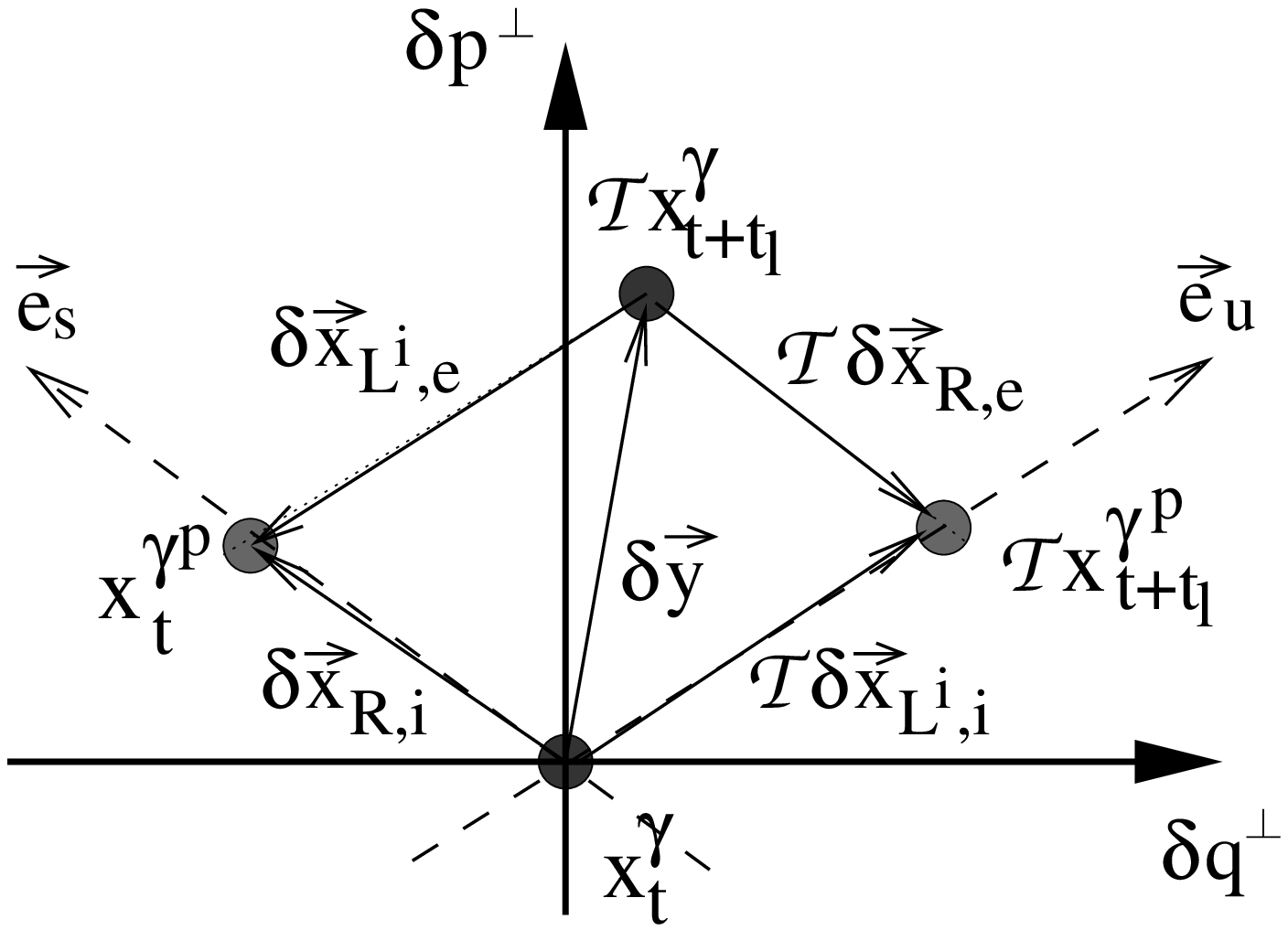}, width=0.4\textwidth}}
\caption{
PSS at ${\bf x}_t$ right
after part ${\cal L}$ and before ${\cal R}$. The displacement
vector $\delta \vec{y}$
points from the orbit $\gamma$ to its time--reversed version
$\gamma^i$. The deviation of the partner orbit
$\gamma^p$ from the original orbit $\gamma$ is
described by the vector $\delta \vec{x}_{R,i}$.
Shown is only a two--dimensional projection of the
$(2 f-2)$-dimensional PSS. 
}
\label{fig:y_in_pss}
\end{figure}

Let us denote by $\delta \vec{x}_{{\cal R},i}$ the phase--space displacement
between $\gamma$
and $\gamma^p$ at the
beginning of part ${\cal R}$ (time $t$),
see Fig.~\ref{fig:y_in_pss}. To simplify the notations, we do not write explicitly  the
dependence of the displacement vectors  on 
${\bf x}^\gamma_{t}$ and $t_{\rm l}$.
At the end of part
${\cal R}$, i.e., at time $t+t_{\rm l}$, the displacement $\delta \vec{x}_{{\cal R},i}$ has
changed to 
$\delta \vec{x}_{{\cal R},e}$.
At the beginning and the end of part ${ \cal L}$, the displacement vectors
between the time--reversed orbit
$\gamma^i$ and the partner orbit $\gamma^p$
are denoted by $\delta \vec{x}_{{\cal L}^i,i}$ 
and $\delta \vec{x}_{{\cal L}^i,e}$, respectively. Here,
${\cal L}^i$ indicates that one has to invert time on ${\cal L}$. 
The vectors $\delta \vec{x}$ are given  explicitly by
\begin{eqnarray} \label{eq-displacement_vec}
\delta \vec{x}_{{\cal R},i}  & = & {\bf x}^{\gamma^p}_t - {\bf x}^\gamma_t
\; , \quad
\delta \vec{x}_{{\cal L}^i,i}  =  {\bf x}^{\gamma^p}_{t+t_{\rm l} } - \T {\bf x}^\gamma_{t}
\; , \; \\
\delta \vec{x}_{{\cal R},e}  & = & {\bf x}^{\gamma^p}_{t+t_{\rm l} } - {\bf x}^\gamma_{t+t_{\rm l}}
\quad \text{and} \quad
\delta \vec{x}_{{\cal L}^i,e}  =  {\bf x}^{\gamma^p}_{t} - \T {\bf x}^\gamma_{t+t_{\rm l} } \;.
\nonumber
\end{eqnarray}
The vectors $\delta \vec{x}_{{\cal R},i}$ and  $\delta \vec{x}_{{\cal L}^i,e}$
lie in the PSS
defined at  ${\bf x}^\gamma_{t}$ (see Fig.~\ref{fig:y_in_pss}), while
$\delta \vec{x}_{{\cal R},e}$ and $\delta \vec{x}_{{\cal L}^i,i}$
are in the PSS at ${\bf x}^\gamma_{t+t_{\rm l}}$. 
Let $R=M(t_{\rm l},{\bf x}_t^\gamma)$ and
$L= M( T_\gamma - t_{\rm l},{\bf x}_{t+t_{\rm l}}^\gamma)$ be the stability matrices of the parts 
${\cal R}$ and ${\cal L}$ of $\gamma$, respectively. 
The stability matrix of ${\cal L}^i$ is given by 
$L^i = \T \, L^{-1} \, \T$.
Since the partner orbit $\gamma^p$ is assumed 
to follow closely $\gamma$ on part ${\cal R}$ and $\gamma^i$
on part ${\cal L}$, one can use the linear approximation to evaluate
$\delta \vec{x}_{{\cal R},e}$ 
and $\delta \vec{x}_{{\cal L}^i,e}$ as functions of $\delta \vec{x}_{{\cal R},i}$
and $\delta \vec{x}_{{\cal L}^i,i}$, respectively, i.e.,
\begin{subequations}
\label{lin_equations_partner}
\begin{eqnarray}
\label{lin_equations_partner_1}
        \delta \vec{x}_{{\cal R},e} = R \, \delta \vec{x}_{{\cal R},i} \quad , \quad
        \delta \vec{x}_{{\cal L}^i,e} =  L^i \, \delta \vec{x}_{{\cal L}^i,i}.
\end{eqnarray}
These equations determine the
two single parts of the partner orbit $\gamma^p$ during ${\cal R}$
and ${\cal L}$.
In addition, the relations
\begin{equation}
\label{lin_equations_partner_2}
        \delta \vec{x}_{{\cal R},i} - \delta \vec{x}_{{\cal L}^i,e}  
         =  \delta \vec{y}  \quad , \quad
        \delta \vec{x}_{{\cal R},e} - \delta \vec{x}_{{\cal L}^i,i} 
         =  -\T\delta \vec{y} \;
\end{equation}
\end{subequations}
make sure that the two parts fit together in the
encounter region. The set of
equations (\ref{lin_equations_partner}) can
be rewritten to give
\begin{equation} \label{eq-x_Ri}
( 1 - L^i R)\, \delta \vec{x}_{{\cal R},i}  =  ( 1 + L^i \T ) \, \delta \vec{y} 
\qquad \text{and} \qquad
(1 - R L^i )\,\delta \vec{x}_{{\cal L}^i,i}  =  ( \T + R ) \,\delta \vec{y} 
\;.
\end{equation}
Assuming that the determinants of $(1 - L^i R)$ and $(1 - R L^i)$ do not vanish,
the system of linear equations (\ref{lin_equations_partner}) has a unique solution.
This solution yields the vectors $\delta \vec{x}$ in terms of the
displacement $\delta \vec{y}$. Hence it
characterizes the geometry of the partner orbit $\gamma^p$ in terms
of deviations from $\gamma$ and $\gamma^i$. 

It is important to note that all points
${\bf x}_{t+\Delta t}^\gamma$ within the encounter 
region lead to the same partner orbit $\gamma^p$. This means that, 
when writing Eqs.~(\ref{lin_equations_partner})
for position ${\bf x}_{t+\Delta t}^\gamma$ instead of ${\bf x}_{t}^\gamma$,
the solution is just the vector $\delta \vec{x}_{{\cal R},i}$  
corresponding to ${\bf x}_t^\gamma$ shifted along the orbit  
during time $\Delta t$ and similarly for the other  vectors
$\delta \vec{x}$ in Eq.~(\ref{eq-displacement_vec}).
To see this, let us first remark that the time
evolution  of $\delta \vec{y}$ between $t$ and $t+\Delta t$ is determined  
by the stability matrix $M = M(\Delta t , {\bf x}_t^\gamma )$ via
$\delta \vec{y}_{\Delta t} = M \delta \vec{y}$.
Similar relations hold for the vectors
$\delta \vec{x}$ in Eq.~(\ref{eq-displacement_vec}).
The linearization of the equation of motion 
is,  by definition, justified within the whole encounter region. 
The replacement of $( {\bf x}_t^\gamma , t_{\rm l})$ 
by $({\bf x}_{t+\Delta t}^\gamma ,t_{\rm l}- 2 \Delta t)$  
thus amounts to the transformations
\begin{equation}
\begin{array}{clclclc}
\label{shift_pss}
        \delta \vec{y} \, & \to & \, M \, \delta \vec{y}
         & \quad , \quad &  & & \\
        \delta \vec{x}_{{\cal R},i} \, & \to & \, M \, \delta \vec{x}_{{\cal R},i} 
        & \quad , \quad &
        \delta \vec{x}_{{\cal R},e} \, & \to & \, ( M' )^{-1} \, \delta \vec{x}_{{\cal R},e} \; ,
          \\
         \delta \vec{x}_{{\cal L}^i,i}  \,  & \to &
             \, ( M^i )^{-1} \, \delta \vec{x}_{{\cal L}^i,i} 
        & \quad , \quad &
       \delta \vec{x}_{{\cal L}^i,e}  \, & \to & \, ( M ' )^i  \, 
        \delta \vec{x}_{{\cal L}^i,e} \; ,
          \\
        R \,  & \to &  \, (M ')^{-1} \, R \, M^{-1} 
        & \quad , \quad &
        L \, & \to & \,  M \, L \, M ' 
        \, ,
\end{array}
\end{equation}
with $M' = M (\Delta t, {\bf x}_{t+t_{\rm l}^\gamma - \Delta t})$.
One can easily check that the set of equations  (\ref{lin_equations_partner}) 
is invariant under these transformations.
This means that the same partner
orbit $\gamma^p$ is obtained
no matter whether ${\bf x}_t^\gamma$ or ${\bf x}_{t+\Delta t}^\gamma$ was chosen 
within the
encounter region.

Let us first restrict our considerations to the case of long  
parts ${\cal R}$ and ${\cal L}$.
This has to be understood in the sense that the linear approximation
with respect to the evolution of $\delta \vec{y}$ breaks down at some
time between $t$ and $t+t_{\rm l}$ and similarly, going backward in time, 
between $t$ and $t+t_{\rm l} -T_\gamma$. This means that
$t_{\rm l} > \Delta t_u$ and $T_\gamma - t_{\rm l} > \Delta t_s$.
We first note that these two conditions actually imply the stronger restriction
\begin{equation} \label{eq-min_max_loop_time}
2 \Delta t_u  < t_{\rm l} < T_\gamma - 2 \Delta t_s \;
\end{equation}
because the displacements $\delta \vec{y}$ 
at the beginning and the end of parts ${\cal R}$ and ${\cal L}$ are related
to each other via the time--reversal operator ${\T}$.
Formally this can be seen as follows. 
The displacement $\delta \vec{y}_{\Delta t} = \delta \vec{y} ( {\bf x}_{t + \Delta t}^\gamma, t_{\rm l}-2 \Delta t)$ 
satisfies
$\delta \vec{y}_{t_{\rm l} - \Delta t} = -\T \delta \vec{y}_{\Delta t}$, as is easily
checked with Eq.~(\ref{delta_y}). 
Let us imagine that $\Delta t_u > t_{\rm l}/2$ implying
that the linear approximation 
$\delta \vec{y}_{\Delta t} = M \delta \vec{y}$ is still
valid after ${\bf x}_{t+\Delta t}^\gamma$ reaches 
the middle of ${\cal R}$. This would imply that
$| \delta \vec{y}_{\Delta t}| $  continues 
to increase exponentially with $\Delta t$
after time $t_{\rm l}/2$ until $\Delta t$ reaches $\Delta t_u$. Such a 
statement is in  contradiction 
with the above--mentioned identity. This shows that $\Delta t_u < t_{\rm l}/2$
must hold. 
A similar argument on part ${\cal L}$ shows the second inequality in 
Eq.~(\ref{eq-min_max_loop_time}). 

For a long part ${\cal R}$ fulfilling (\ref{eq-min_max_loop_time}),
the stability matrix $R$ in Eq.~(\ref{lin_equations_partner_1}) is 
characterized by exponentially
large stretching factors $\Lambda_i({\bf x}_t^\gamma, t_{\rm l})$.
Substituting Eq.~(\ref{delta_y}) into Eq.~(\ref{eq-x_Ri}) 
and  using Eq.~(\ref{eq-time_reverse_e^u,s}),
one thus finds
\begin{equation}
\begin{array}{lllll}
\label{slep_approx}
        \delta \vec{x}_{{\cal R},i}   & = & \delta \vec{y}_s 
         & = & \sum\limits_{i=1}^{f-1} s_i \; 
        \vec{e}^{\, s}_i( {\bf x}^\gamma_{t}) \; ,
         \\
        \delta \vec{x}_{{\cal L}^i,i} & = & \T  \delta \vec{y}_u 
        & = & \sum\limits_{i=1}^{f-1} u_i \;
        \vec{e}^{\, s}_i( \T {\bf x}^\gamma_{t})  \; ,
         \\
        \delta \vec{x}_{{\cal R},e} & = & 
          - \T  \delta \vec{y}_s 
        & = &  - \sum\limits_{i=1}^{f-1} s_i \;
        \vec{e}^{\, u}_i(\T {\bf x}^\gamma_{t}) \; ,
         \\
        \delta \vec{x}_{{\cal L}^i,e} & = & - \delta \vec{y}_u 
        & = & - \sum\limits_{i=1}^{f-1} u_i \;
        \vec{e}^{\, u}_i({\bf x}^\gamma_{t}) \; .
\end{array}
\end{equation}
This solution is correct up to first order in the
small quantities $s_i$ and $u_i$. Terms smaller than $s_i$ and $u_i$ by a factor
$e^{-t_{\rm l} \lambda_i^\gamma}$ or $e^{-(T_\gamma - t_{\rm l} )\lambda_i^\gamma}$ have been also neglected.
It means that due to the large lengths of both parts ${\cal R}$
and ${\cal L}$ the vectors $\delta \vec{x}_{{\cal R},i}$ and $\T \delta \vec{x}_{{\cal L}^i,i}$
describing the
partner orbit have to lie very close to the stable and the unstable manifolds  
at ${\bf x}_t^\gamma$, respectively~\cite{Braun02b}.
Furthermore, the points 
${\bf x}_t^\gamma$, ${\bf x}_{t}^{\gamma^p}$, $\T{\bf x}_{t+t_{\rm l}}^\gamma$, and 
$\T{\bf x}_{t+t_{\rm l}}^{\gamma^p}$ form a parallelogram in phase 
space~\cite{Spehner03,Turek03}, see Fig.~\ref{fig:y_in_pss}.

It is important to notice that there can be a small
set of vectors $\delta \vec{y}$
for which (\ref{eq-min_max_loop_time}) does not hold. 
This is the case when
either of the parts, say ${\cal R}$, is too short~\cite{Spehner03,Turek03,Mueller03}.
Then the orbit $\gamma$ and the time--reversed orbit $\gamma^i$
stay close together inside the whole part ${\cal R}$ so that  ${\cal R}$ is contained
within the encounter region.
This means that ${\cal R}$ is an almost self--retracing part of
trajectory in configuration
space. This may happen,  
for example, in billiards with  hard walls
if one of the reflections is almost perpendicular to the 
boundary~\cite{Mueller01, Mueller03}.
If there is no potential or hard wall, 
as in the case of the geodesic flow on
a Riemann surface with constant negative curvature~\cite{Sieber01,Sieber02},
trajectories with almost self--retracing parts cannot exist.
If ${\cal R}$ is contained within the encounter region, 
the linear approximation $\delta \vec{y}_{\Delta t}  =
 M( \Delta t,{\bf x}_t^\gamma) \delta \vec{y}$ can be 
 applied to
$\delta \vec{y} = \delta \vec{y}({\bf x}_t^\gamma , t_{\rm l})$
at least up to $\Delta t = t_{\rm l}$. This leads to the additional equation 
$R \, \delta \vec{y} = -\T \delta \vec{y}$ besides Eq.~(\ref{lin_equations_partner}). For indeed,
following the linearized motion around ${\cal R}$, we see that ${\bf x}_t$ and 
$\T{\bf x}_{t+t_{\rm l}}$ are interchanged and reverted in time.
The solution (\ref{eq-x_Ri}) is then $\delta \vec{x}_{{\cal R},i} = \delta \vec{y}$,
$\delta \vec{x}_{{\cal R},e} = - \T \delta \vec{y}$ and 
$\delta \vec{x}_{{\cal L}^i,i} = \delta \vec{x}_{{\cal L}^i,e} = \vec{0}$.
This means that if the time $t_{\rm l}$ violates the
condition (\ref{eq-min_max_loop_time}), 
the solution of Eq.~(\ref{lin_equations_partner}) does not
yield a new partner orbit but just the time--reversed orbit $\gamma^i$. 
Since the orbit pairs $(\gamma, \gamma^i)$ are 
already accounted for in the diagonal approximation
(\ref{K_diag})  one must only consider
intersection points $\delta \vec{y}({\bf x}_t^\gamma ; t_{\rm l})$
in the PSS which fulfill (\ref{eq-min_max_loop_time}). In other words,
the length $t_{\rm l}$ of part ${\cal R}$ must be large
enough so that the linear approximation for $\delta \vec{y}_{\Delta t}$
breaks down for $\Delta t < t_{\rm l}$ and
similarly for part ${\cal L}$.
Note that $\Delta t_s$ and $\Delta t_u$ are large 
(of order $\lambda_j^{-1} \ln |s_j^{-1}|$
and $\lambda_k^{-1} \ln |u_k^{-1}|$, respectively,
see Subsection~\ref{subsec:encounter}) if the components $\{ s_i,u_i \}$ 
of $\delta \vec{ y}$ are small.

\subsection{Action difference, orbit weights, and Maslov indices}
\label{subsec:action_difference}

The action difference $\Delta S$
between the orbit $\gamma$ and its partner orbit
$\gamma^p$ can be found by expanding
the action of $\gamma^p$ in part ${\cal R}$
in terms of $\gamma$ and in part ${\cal L}$ in terms
of $\gamma^i$. The derivation of $\Delta S$ is the same
as for systems with $f=2$ degrees of
freedom~\cite{Spehner03,Turek03}.
By using the parallelogram property (\ref{slep_approx}),
which is justified 
since $t_{\rm l}> 2 \Delta t_u$ and $T_\gamma - t_{\rm l}> 2 \Delta t_s$ 
are large, one finds that $\Delta S$ is given
in terms of the components $\{ s_i, u_i\}$ of the
displacement $\delta \vec{y}({\bf x}_t^\gamma, t_{\rm l})$ by
\begin{equation}
\label{action_diff}
  \Delta S \equiv S_\gamma - S_{\gamma^p} 
  \approx \delta \vec{y}_u^{\, T} \, \Sigma \, \delta \vec{y}_s
  = \sum\limits_{j=1}^{f-1} S_{\text{cl}} s_j u_j \equiv \sum\limits_{j=1}^{f-1} S_j \, .
\end{equation}
Thus $\Delta S$ equals the symplectic area of the parallelogram introduced 
in Subsection~\ref{subsec:partner_orbits}, see Fig.~\ref{fig:y_in_pss}. 
In the last two equalities in Eq.~(\ref{action_diff}), 
we have used Eqs.~(\ref{units}, \ref{delta_y}) 
and defined $S_j\equiv S_{\text{cl}}s_j u_j$.
The approximation (\ref{action_diff}) is correct up to second order in the
small $|s_i|, |u_i| \ll 1$. It is consistent
with the concept of the encounter region as it
yields the same action difference no matter at what position 
${\bf x}_{t + \Delta t}^\gamma$
within the encounter region it is evaluated. This is
due to the conservation of the symplectic product
under the dynamics. As only small action differences
$\Delta S \sim \hbar$ contribute significantly
to the semiclassical form factor (\ref{gen_K}),
the restriction of the considerations presented above
to small components $|s_i|,|u_i| \sim \sqrt{\hbar / S_{\rm cl}} \ll 1$
is well justified.

Besides the two different actions $S_\gamma$ and $S_{\gamma^p}$
entering the semiclassical form factor (\ref{gen_K}), one must also compare
the weights $w_\gamma$ and $w_{\gamma^p}$ given by Eq.~(\ref{eq-w}).
These weights are equal up to small corrections
of first order in $u_i$ and $s_i$ as can be shown
in the following way. First of all, for any continuous function $f({\bf x})$
defined in phase space one finds, using Eq.~(\ref{partner_coords}),
\begin{equation}
\label{f_along_partner}
        \int\limits_0^{T_{\gamma^p}} \diff t' \,
        f({\bf x}^{\gamma^p}_{t'}) \simeq
        \int\limits_t^{t+t_{\rm l}} \diff t' \,
        f({\bf x}^{\gamma}_{t'}) +
        \int\limits_{t+t_{\rm l}}^{T_\gamma + t} \diff t' \,
        f({\cal T}{\bf x}^{\gamma}_{2t + t_{\rm l} - t'})
\end{equation}
with small corrections of the order of
$|s_i|,|u_i| \sim \sqrt{\hbar / S_{\rm cl}}$.
That means that the integral over any function
$f({\bf x})$ along the partner orbit $\gamma^p$ is
approximately given by integrals along parts of $\gamma$
and $\gamma^i$.
The corrections in Eq.~(\ref{f_along_partner})
are primarily due to the deviations of the
partner orbit $\gamma^p$ from the original orbit $\gamma$
or its time--reversed version $\gamma^i$ within the
encounter region. Obviously, Eq.~(\ref{f_along_partner})
yields $T_\gamma \simeq T_{\gamma^p}$ for $f({\bf x})=1$.
Similarly, we can apply Eq.~(\ref{f_along_partner}) to the
local growth rates $f({\bf x}) = \chi_i({\bf x})$,
which results into $\lambda_i^\gamma \simeq \lambda_i^{\gamma^p}$
in view of Eq.~(\ref{eq-Lyapunov_exp}) and of the identity 
$\chi_i (\T {\bf x}) = \chi_i ({\bf x})$. 
Hence the Lyapunov exponents of the two partner orbits $\gamma$ and $\gamma^p$ have
to be almost equal. Finally, we can also
identify $f({\bf x})$ with the local change in the
winding number of the stable or unstable manifolds
which allows for a calculation of
the Maslov indices~\cite{Foxman95}.
As the winding number of a periodic orbit 
has to be an integer one finds that, for smooth dynamics, the Maslov
index of the partner orbit has to be exactly equal to the
Maslov index of the original orbit~\cite{Turek03,Mueller03}, i.e., $\mu_\gamma = \mu_{\gamma^p}$.
Putting these results together in Eq.~(\ref{eq-w}), one concludes that
$w_\gamma \simeq w_{\gamma^p}$.  
In the spirit of a stationary phase approximation
we therefore keep only the action difference 
$\Delta S = S_{\gamma} - S_{\gamma^p}$ in the phase while neglecting
small differences in the pre-exponential factors in
Eq.~(\ref{gen_K}).

\subsection{Statistics of partner orbits and the spectral form factor}
\label{subsec:statistics_of}

In the following we show how the
orbit pairs $(\gamma, \gamma^p)$ specified above
determine the next-to-leading-order result
for the spectral form factor. We assume that the
dominant terms beyond the diagonal approximation
in Eq.~(\ref{gen_K}) are due to the systematic
action correlations of these
orbit pairs.
Thus the double sum over periodic orbits (\ref{gen_K}) can be replaced 
by a single sum over the orbits $\gamma$
followed by a sum over
all the partner orbits $\gamma^p$ of $\gamma$ while
all other terms are neglected, i.e.,
\begin{equation}
\label{eq-spect_K_2}
K^{(2)}(T) = \frac{T}{T_H} \,
 \left\langle \left\langle 
   \sum_{\text{partners $\gamma^p$}} 
    \exp \left( \I \frac{S_\gamma - S_{\gamma^p}}{\hbar} \right) 
  \right\rangle_{\text{po},T} \right\rangle_{\Delta E}
\end{equation}
where the periodic--orbit average over $\gamma$ is given by Eq.~(\ref{po_average}).
All partner orbits $\gamma^p$ of  $\gamma$ are
characterized by the set of action differences $\{S_j\}$ defined
in Eq.~(\ref{action_diff}). Therefore,  setting $\tau=T/T_H$, the
sum over the partner orbits in 
Eq.~(\ref{eq-spect_K_2}) can be rewritten as an integral over the $S_j$'s,
\begin{widetext}
\begin{equation}
\label{K2_partners}
 K^{(2)}( \tau ) = \tau \left\langle \int\limits_{-S_{\rm max}(E)}^{S_{\rm max}(E)}
 {\rm d}S_1 \dots {\rm d}S_{f-1} \, 
 \left\langle \frac{{\rm d}^{f-1} N_\gamma(\{S_j\})}
                   {{\rm d}S_1 \dots {\rm d}S_{f-1}}
 \right\rangle_{{\rm po}, \tau T_H}
 \exp \left( {\rm i}
  \sum\limits_{j=1}^{f-1} \frac{S_j}{\hbar} \right) \right\rangle_{\Delta E} \, ,
\end{equation}
\end{widetext}
where $S_{\rm max}$ stands for the maximal action difference
occurring among the pairs of partner orbits.
The density of partner orbits $\gamma^p$ for a given orbit
$\gamma$ with the set of action differences $\{S_j\}$ is denoted by
${\rm d}^{f-1} N_\gamma (\{S_j\}) / {\rm d}S_1 \dots {\rm d}S_{f-1}$. This quantity
is the crucial ingredient, and we will show how its periodic--orbit average  can
be calculated in ergodic systems with an arbitrary number of degrees of freedom.
In contrast to the case of two--dimensional systems, the
derivation is significantly more involved because of the
higher number of stable and unstable coordinates, Lyapunov exponents,
and the maximum condition (\ref{t_su}).

\begin{figure}
\centerline{
\epsfxsize=0.40\textwidth
\epsfbox{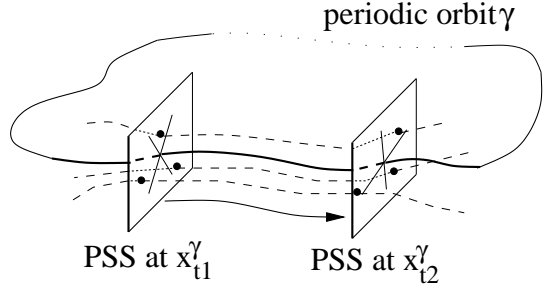}}
\caption{
\label{fig:pss_along_po}
Sketch of the PSS as it is shifted along
the periodic orbit $\gamma$ (solid line).
 Three pieces of the time--reversed orbit
$\gamma^i$ are represented by dashed lines.
If the PSS
moves with the flow in phase space from ${\bf x}_{t1}^\gamma$
to ${\bf x}_{t2}^\gamma$ all intersection points of the PSS
with  $\gamma^i$ change their
positions according to
the linearized equations of motion (\ref{lin_motion}). Note that
not only $\gamma^i$ but also
$\gamma$ itself could come close to ${\bf x}_{t1}^\gamma$ at a
later time. However, we have not include this in the sketch above.}
\end{figure}

Let us for a moment fix one point 
${\bf x}$ on $\gamma$ (to simplify the notation,
we omit here the superscript $\gamma$ on ${\bf x}$ and choose temporarily the origin 
of time such that ${\bf x}_t = {\bf x}$ for $t=0$), and consider the PSS 
${\mathcal P}$ perpendicular to the orbit at  ${\bf x}$. The time--reversed 
orbit $\gamma^i$ pierces through ${\mathcal P}$ many 
times. Some of these piercings~---~each of it associated with a different
time $t_{\rm l}$~---~occur at points 
${\cal T}{\bf x}_{t_{\rm l}}$ close to ${\bf x}$,
see Figs.~\ref{fig:encounter_region} and \ref{fig:pss_along_po}. 
Let $\rho_\gamma (\{s_i,u_i\}; {\bf x}) \,\D^{f-1} s \, \D^{f-1} u$ 
denote the number of such intersection points, with stable and unstable components 
of $\delta\vec{y}= {\cal T}{\bf x}_{t_{\rm l}}-{\bf x}$
lying in intervals $(s_i,s_i+ \D s_i)$ and 
$(u_i,u_i+ \D u_i)$, respectively. We exclude from $\rho_\gamma$ all points 
${\cal T}{\bf x}_{t_{\rm l}}$ violating 
the condition $2\Delta t_u<t_{\rm l} <T_\gamma-2 \Delta t_s$
since they either do not exist at all or
do not give rise to a distinct partner orbit. Thus we have 
the density of valid intersection points,
\begin{widetext}
\begin{equation}
\label{def:density}
 \rho_\gamma (\{s_i, u_i\}; {\bf x}) = 
        \int\limits_{2 \Delta t_u}^{T_\gamma-
         2 \Delta t_s} \diff t_{\rm l} \,
        \delta\left([{\bf x} - \T {\bf x}_{t_{\rm l}}]_\parallel \right)
        \prod\limits_{i=1}^{f-1} 
        \delta\left([{\bf x} - \T {\bf x}_{t_{\rm l}}]_{u,i} - u_i \right)
        \delta\left([{\bf x} - \T {\bf x}_{t_{\rm l}}]_{s,i} - s_i \right).
\end{equation}
\end{widetext}
Here, the first delta function ensures that ${\cal T}{\bf x}_{t_{\rm l}}$ lies
in ${\mathcal P}$, i.e., that the coordinate $[{\bf x} - \T {\bf x}_{t_{\rm l}}]_\parallel $ 
of ${\bf x} - \T {\bf x}_{t_{\rm l}}$ in the direction parallel to the orbit 
vanishes.
The lower indices $s$ and $u$  indicate the stable
and unstable components 
of the vectors inside the square brackets.

\begin{figure}
\centerline{
\epsfxsize=0.35\textwidth
\epsfbox{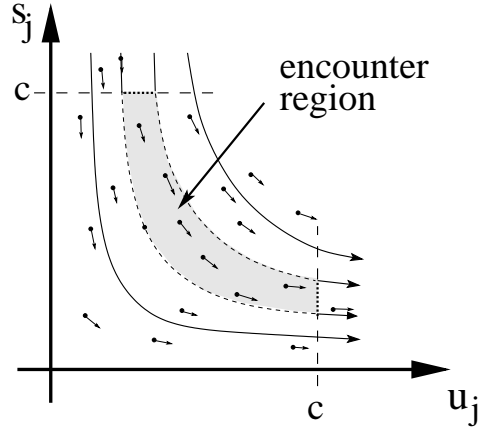}}
\caption{
\label{fig:flow_volume}
Schematic drawing of a projection
of the Poincar{\'e} surface of section (PSS).
The flow of intersection points (black filled circles) is
represented by the thin arrows. For long enough time the unstable
component $u_j$ grows while the stable component $s_j$ shrinks
with $S_j=S_{\text{cl}}s_j u_j$ being constant. There are two ways to
count the intersection points. Either the flux through the $u_j=c$
surface (dotted line) is considered, as in Eq.~(\ref{dN_flow}),
or one counts the number of points in the volume
of the encounter region (dashed area) and  normalizes
that by the time each point
spends in there, as in Eq.~(\ref{dN_volume}).}
\end{figure}

In order to determine how many partner orbits $\gamma^p$ 
of a given fixed orbit $\gamma$ exist with
a given set of action differences $\{S_i\}$,
one has to count the number of corresponding encounter regions of $\gamma$.
As explained in the beginning of this section
each of these encounter regions can be associated to 
a displacement vector $\delta \vec{y}$ or to its corresponding  time--evolved 
$\delta \vec{y}_{\Delta t}$, $-\Delta t_s \leq \Delta t \leq \Delta t_u$.
Therefore we consider the dynamics {\em within}
the PSS 
${\cal P}_t$ at ${\bf x}_t^\gamma$. It can be
parametrized by means of the stable and
unstable coordinates of the different vectors 
$\delta \vec{y} ( {\bf x}_{t}^\gamma , t_{\text{l}} )$
associated to different times $t_{\text{l}}$.
As the PSS is shifted following the phase--space flow 
along the orbit $\gamma$,
the stable and unstable coordinates of each such vector change leaving only 
the products $S_j = S_{\rm cl} \, s_j u_j$ invariant. The vector
$\delta \vec{y}_{\Delta t} = \delta \vec{y}_{\Delta t} ( {\bf x}_{t} , t_{\text{l}} )
\in {\cal P}_{t+\Delta t}$ 
corresponding to a fixed $t_{\text{l}}$
thus moves, as  ${\cal P}_{t+\Delta t}$ is shifted by increasing $\Delta t$,
on a hyperbola as long as ${\bf x}_{t + \Delta t}^\gamma$
remains within the encounter region, i.e., for 
$-\Delta t_s \leq \Delta t \leq \Delta t_u$, see Fig.~\ref{fig:flow_volume}.
Since the number of partner orbits is equal to
the number of encounter regions 
one has now to count each encounter region exactly once.
This can be achieved in two alternative ways:

\begin{enumerate}

\item One can measure
the flux of vectors $\delta \vec{y}$ 
through the hypersurface 
defining the end of the encounter region (see Fig. \ref{fig:flow_volume}). 
According to
the  definition of the encounter region given in Subsection~\ref{subsec:encounter},
 this hypersurface consists of the faces 
$\partial\, {\cal C}_{j}^{\pm} = \{ (s_i, u_i) :  |s_i| \leq c,   |u_{i}| \leq c , u_j = \pm c \}$,
$j=1, \ldots, f-1$,
of the hypercube ${\cal C} = \{ (s_i, u_i) :  |s_i| \leq c,   |u_{i}| \leq c\}$.
The union of all these faces defines a $(2f-3)$-dimensional closed hypersurface 
$\partial\, {\cal C}$ 
contained in the
$(2f-2)$-dimensional PSS. 
The corresponding flux is obtained by multiplying the density $\rho_\gamma$ with the 
component $\dot{u}_j$ of the velocity of the vector
$\delta \vec{y}$ in the   direction  normal
to $\partial\, {\cal C}_{j}^{\pm}$. This velocity 
 is given by $\dot{u}_j = \chi_j({\bf x}_t^\gamma) \,c$,
see Eq.~(\ref{chi}).
Integrating along the orbit we obtain
\begin{widetext}
\begin{eqnarray}
\label{dN_flow}
 \frac{{\rm d}^{f-1} N_\gamma(\{S_i\})}{{\rm d}S_1 \dots {\rm d}S_{f-1}} & = &
        \int\limits_0^{T_\gamma} {\rm d} t \;
        \sum\limits_{j=1}^{f-1} \;
        \int\limits_{-c}^c {\rm d}^{f-1} s \; {\rm d}^{f-1} u \;
        \rho_\gamma (\{s_i, u_i\}; {\bf x}^\gamma_t) \;
        \chi_j ( {\bf x}^\gamma_t) \,c \nonumber \\
        & \; & \qquad 
           \times  \Bigl( \delta (u_j - c) + \delta (u_j + c) \Bigr) \; 
            \left( \prod\limits_{i=1}^{f-1} \delta (S_{\text{cl}} s_i u_i - S_i) \right).
\end{eqnarray}
\end{widetext}
The last product of delta functions restricts the action differences to
the values $\{S_i\}$. 
It should be noted that, since the local growth rate $\chi_j({\bf x}_t^\gamma)$ can
take negative values for some times $t$, the vector 
$\delta \vec{y}$ may also re-enter into the 
hypercube ${\cal C}$ through some  face 
$\partial\,{\cal C}_{k}^{\pm}$  with a negative normal velocity 
$\dot{u}_k  = \chi_k({\bf x}_{t}^\gamma) \,c$ (with possibly $k \not= j$). However, since 
$\delta \vec{y}$ increases exponentially with time at large times, there is one more
passing of $\delta \vec{y}$ through $\partial \,{\cal C}$ 
in the outwards direction ($\dot{u} >0$)
than in the inwards direction ($\dot{u} <0$). The contributions of all subsequent 
passings then mutually cancel each other in Eq.~(\ref{dN_flow}). Hence, for each encounter region, 
only the first crossing of $\partial\,{\cal C}$ at time 
$\Delta t = \Delta t_u$ is accounted for, as required.
Let us also mention that if we had taken any other closed hypersurface 
contained in the hypercube ${\cal C}$ instead of 
$\partial\, {\cal C}$ the same result would have been obtained.
This is because the dynamics 
conserves the number of points in phase space and
thus the number of vectors $\delta \vec{y}$. 

\item 
An alternative version of Eq.~(\ref{dN_flow}), treating all points within the encounter region 
on equal footing, can be found as follows. 
Every vector $\delta \vec{y}$ is counted as long as it remains
within the hypercube ${\cal C}$.
Therefore one has to include the
additional factor of $1/t_{\rm enc}$, since per definition  (\ref{t_enc}), $t_{\rm enc}$ is 
approximately the time
each vector $\delta \vec{y}$ spends within that hypercube. 
The density of partner orbits (\ref{dN_flow})
 can thus be rewritten as
\begin{widetext}
\begin{equation}
\label{dN_volume}
 \frac{{\rm d}^{f-1} N_\gamma(\{S_i\})}{{\rm d}S_1 \dots {\rm d}S_{f-1}} \simeq
        \int\limits_0^{T_\gamma} {\rm d} t \;
        \int\limits_{-c}^c {\rm d}^{f-1} s \, {\rm d}^{f-1} u \,
        \frac{\rho_\gamma (\{s_i, u_i\}; {\bf x}^\gamma_t)}{
              t_{\rm enc}(\{s_i, u_i\}; {\bf x}^\gamma_t)}
        \left( \prod\limits_{i=1}^{f-1} \delta (S_{\text{cl}} s_i u_i - S_i) \right).
\end{equation}
\end{widetext}
More precisely, this expression can be derived as follows 
(also see the Appendix~\ref{app:proof}).
Consider first only the contribution of the encounters at ${\bf x}_t^\gamma$ after time
$t_{\rm l} - 2t$, for a fixed $t_{\rm l}$ and an arbitrary time $t$.
The time duration $t_{\text{enc}}$ of the encounter
and the product $u_i s_i$ of the stable and unstable components
of the vector $\delta \vec{y} ( {\bf x}_{t}^\gamma , t_{\rm l} - 2 t)$ are independent of $t$
as long as ${\bf x}_t^\gamma$ stays within an encounter region,
i.e., while $s_i$ and $u_i$ vary
within $(-c,c)$. 
The time spent by the point characterized by
$\delta \vec{y} ( {\bf x}_{t}^\gamma , t_{\rm l} - 2 t)$
within the hypercube ${\cal C}$ is approximately equal to $t_{\text{enc}}$
in the limit $|u_i|, |s_i| \sim \sqrt{\hbar/S_{\text{cl}}} \ll 1$, where $t_{\text{enc}}$
is large (of order $\ln \hbar$). Indeed, although possible re-entrances of $\delta \vec{y}$ into
${\cal C}$ (see above)
may increase the total time spent by $\delta \vec{y}$ inside ${\cal C}$ to a value
greater than $t_{\text{enc}}$, the relative error made by approximating
it by $t_{\text{enc}}$ is small. 
Using also the fact that the above--specified encounter regions are 
disjoint (if they were overlapping they would define one bigger region),
it follows that the r.h.s. of Eq.~(\ref{dN_volume}) gives
approximately the density of those encounter regions with respect to the action differences
$ \{ S_i \}$. The result (\ref{dN_volume}) then follows by integrating $t_{\rm l}$ over all 
possible values.     

\end{enumerate} 

Actually, expressions (\ref{dN_flow}) and (\ref{dN_volume}) are equivalent. 
Using the fact that the number of vectors $\delta \vec{y}$ is conserved
by the dynamics, 
one can transform the integrals over the hypersurface $\partial\,{\cal C}$
 into an integral over the entire volume of the hypercube. 
More details on this  proof of
the equality between Eq.~(\ref{dN_flow}) and Eq.~(\ref{dN_volume})
are given  in the Appendix~\ref{app:proof}. It is important for what follows 
to note that this equality still holds if the range of integration of 
$t_{\rm l}$ in Eq.~(\ref{def:density})
is replaced by the larger interval  $(0, T_\gamma)$ which
corresponds to the additional inclusion of intersection
points $\delta \vec{y}$ that cannot be associated with
a partner orbit. The reason for
the introduction of the two different expressions for
the number of partner orbits is because of its crucial
importance from a technical point of view. We will apply
either Eq.~(\ref{dN_flow}) or Eq.~(\ref{dN_volume}) depending
on which one can be calculated easier. It will turn out
that this allows for a major simplification of the
derivations to follow. In particular, the complicated
analytic structure of $t_{\text{enc}}(\{s_i,u_i\};{\bf x})$,
see Eqs.~(\ref{t_su}) and  (\ref{t_enc}),
does not directly enter the calculations.

The periodic--orbit  average of the 
density of partner orbits
in  Eq.~(\ref{dN_flow}) or Eq.~(\ref{dN_volume}) can be transformed into an average over the 
constant--energy surface  by means of the sum rule 
(\ref{sum_rule}). After this step has been performed,  the density 
$\rho_\gamma$ has to be evaluated 
at arbitrary  points ${\bf x}$ on a set of measure one inside the 
constant--energy surface,
instead of taking the points ${\bf x}_t^\gamma$ belonging to periodic orbits
as the arguments.
For such points ${\bf x}$ one can neglect 
classical correlations between ${\bf x}$ and
${\cal T}{\bf x}_{t_{\rm l}}$ for
$2 \Delta t_u \leq t_{\rm l} \leq T - 2 \Delta t_s$. This is because
$\lambda_i^{-1} \ll \Delta t_{u,s} \ll T$ 
in the relevant limit $|s_i|,|u_i| \sim \sqrt{\hbar / S_{\rm cl}} \ll 1$, 
$T \sim T_H \sim \hbar^{1-f}$.
More precisely,  ergodicity allows one to 
approximate  the time integral in  Eq.~(\ref{def:density}) 
by a phase--space average,
\begin{widetext}
\begin{eqnarray}
\label{ergodic:density}
\nonumber
\rho (\{ s_i, u_i \}; {\bf x} ) 
& \approx &
\frac{T - 2 \Delta t_s (\{ s_i\}; {\bf x} ) - 2 \Delta t_u (\{  u_i \} ; {\bf x} )}{\Omega} 
 \int \D {\bf x}' \, \delta ( E - H ({\bf x}' ) )\,
  \delta\left([{\bf x} - \T {\bf x}' ]_\parallel \right)
\\
& &
        \prod\limits_{i=1}^{f-1} 
        \delta\left([{\bf x} - \T {\bf x}' ]_{u,i} - u_i \right)
        \delta\left([{\bf x} - \T {\bf x}' ]_{s,i} - s_i \right)\;.
\end{eqnarray}
\end{widetext}
Since the Jacobian of the transformation 
$( \delta {\bf q}^\perp , \delta {\bf p}^\perp ) \to ( s_i, u_i )$
gives a factor $S_{\text{cl}}^{f-1}$ this yields  
\begin{eqnarray}
\label{result:density}
        \rho(\{s_i, u_i\}; {\bf x}) & \approx &
        \rho^{\rm lead} + \rho^{\rm corr}(\{s_i, u_i\}; {\bf x})
        \\
        & = &
        \frac{S^{f-1}_{\rm cl}}{\Omega}  T - 
        \frac{S^{f-1}_{\rm cl}}{\Omega}\, 2 t_{\rm enc}(\{s_i, u_i\}; {\bf x}) \; .
\nonumber
\end{eqnarray}
Therefore $\rho$ is given by a leading
contribution plus a small  correction term due to the exclusion of short 
times $t_l$ violating condition~(\ref{eq-min_max_loop_time}). 
The corrections to the ergodic approximation are not written in 
Eqs.~(\ref{ergodic:density}) and (\ref{result:density}). Although they may be bigger than 
$\rho^{\text{corr}}$, one expects 
them to be strongly reduced after averaging ${\bf x}$ over the constant--energy
surface, as required by the sum rule. 
This is not the case for the correction term $\rho^{\text{corr}}$, which, 
as we shall see now, determines the form factor.

Indeed, if only the leading term $\rho^{\rm lead}$ in the density (\ref{result:density})
is considered,   
one finds that the form factor (\ref{K2_partners})
vanishes in the semiclassical limit for the following reason.
As $\rho^{\rm lead}$ does not depend on ${\bf x}$ and $\{s_i, u_i\}$, 
its contribution to the density of partner orbits can be most easily calculated by means of
Eq.~(\ref{dN_flow}). It yields
\begin{equation}
\label{result:dN_lead}
        \left\langle \frac{{\rm d}^{f-1} N_\gamma(\{S_i\})}
           {{\rm d}S_1 \dots {\rm d}S_{f-1}}
        \right\rangle_{{\rm po}, T}^{\rm (lead)} \approx
        2^{f-1} \; \frac{T^2}{\Omega} \; \sum\limits_{j=1}^{f-1}
        \lambda_j \; \prod\limits_{i\neq j}^{f-1} \ln \left( \frac{S_{\text{cl}} c^2}{|S_i|} 
         \right).
\end{equation}
Here we have used the identity $\overline{\chi_j({\bf x})} = \lambda_j$.
If this result (\ref{result:dN_lead})
is inserted into the expression for the form factor (\ref{K2_partners}),
one obtains $K^{(2)}(\tau) = 0$ due to the energy average. 

Therefore the small correction term $\rho^{\rm corr}$ given in 
Eq.~(\ref{result:density}) is of crucial importance. 
To determine its contribution to the form factor, it turns out to be technically
favorable to use expression (\ref{dN_volume})
instead of Eq.~(\ref{dN_flow}) for the
density of partner orbits. The reason is that  the two appearances in 
Eqs.~(\ref{dN_volume}) and (\ref{result:density}) of $t_{\rm enc}$ mutually cancel. 
Inserting $\rho^{\rm corr}$ from Eq.~(\ref{result:density})
into Eq.~(\ref{dN_volume}), one finds
\begin{equation}
\label{result:dN_corr}
        \left\langle \frac{{\rm d}^{f-1} N_\gamma(\{S_i\})}{{\rm d}S_1 \dots {\rm d}S_{f-1}}
        \right\rangle_{{\rm po}, T}^{\rm (corr)} =
        - 2^{f} \, \frac{T}{\Omega} \,
        \prod\limits_{i=1}^{f-1} \ln \left( \frac{S_{\text{cl}} c^2}{|S_i|} \right) \;.
\end{equation}

The result (\ref{result:dN_lead}) together with Eq.~(\ref{result:dN_corr})
gives the correct asymptotic form of 
the averaged density of partners in the limit $\hbar \to 0$, $\tau=T/T_H$ fixed. 
 Since the leading term (\ref{result:dN_lead}) gives a vanishing contribution
to the form factor (\ref{K2_partners}), only the correction (\ref{result:dN_corr}) 
determines the final result,
\begin{widetext}
\begin{equation}
\label{result:K2}
        K^{(2)}(\tau) \simeq -2 \tau^2 \frac{T_H}{\Omega}
        \prod\limits_{i=1}^{f-1} \left\{ - 2 
           \int\limits_{-\infty}^{\infty}
        {\rm d} S_i \, \exp \left( {\rm i} \frac{S_i}{\hbar} \right) \,
        \ln \left( \frac{|S_i|}{S_{\text{cl}} c^2} \right) \right\} \simeq - 2 \tau^2 \, .
\end{equation}
\end{widetext}
This result is universal, i.e., it does not contain any information about the set
of Lyapunov exponents $\{ \lambda_i \}$ or the constant $c$ defining the encounter
region. Thus as our first major result we find that the next-to-leading
order correction beyond the diagonal approximation agrees 
with the BGS conjecture independently of the
number of degrees of freedom the system possesses.

\section{Matrix elements fluctuations}
\label{sec:matrix-element_fluct}

The aim of this section is to evaluate the generalized
form factor $K_{ab} (T)$ defined in Eq.~(\ref{eq-form_factor}) based on the method developed in the previous
section.
This form factor describes the correlations of the diagonal 
matrix elements $\langle n | \hat{a} | n \rangle$ and $\langle m | \hat{b} | m \rangle$,
corresponding to distinct energies $E_n$ and $E_m$,
of two given quantum observables $\hat{a}$ and $\hat{b}$. We assume that 
$\hat{a}$ and $\hat{b}$ have well--behaved 
classical limits given by smooth Weyl symbols $a({\bf x})$ and $b({\bf x})$.

\subsection{Leading term}
\label{the_leading_term}

To zeroth order in $\hbar$, the form factor (\ref{eq-form_factor}) is given by  
\begin{equation} \label{eq-leadingterm}
K_{ab} (T) \approx  \overline{a ({\bf x})}\; \overline{b ({\bf x})} \, K (T) \;.
\end{equation}
Actually,
Snirelman's theorem~\cite{Snirelman74} for classical ergodic flows   
implies that the Wigner functions  of almost all 
eigenstates $| n \rangle$ with energies $E_n$ converging to $E$ are 
uniformly distributed over the constant--energy surface $H ({\bf x} ) = E$ in the semiclassical limit.
Equivalently, this means that
the matrix elements $\langle n | \hat{a} | n \rangle$ 
converge to the average (\ref{def:ps_average}),
\begin{eqnarray} \label{eq-Snirelman}
& \langle n | \hat{a} | n \rangle \to \overline{a ({\bf x})} \nonumber \\
& \text{as} \quad \hbar \to 0 \; , \quad n \to \infty
\quad \text{such that} \quad E_n \approx E \; .
\end{eqnarray}
It is worthwhile to mention that this is   only true  for eigenstates $| n \rangle$ 
of the quantum Hamiltonian pertaining to a ``set of density one''
\footnote{
More precisely, the number
of eigenstates satisfying Eq.~(\ref{eq-Snirelman})
with eigenenergies $E_n$ in $[E-\Delta E, E + \Delta E]$, 
divided by the total number of eigenstates with eigenenergies in this interval,  
tends to $1$ in the limits $\hbar \to 0$, 
$\langle {d}(E) \rangle^{-1}_{\Delta E} \ll \Delta E \ll E$.
}.
Heller's scars~\cite{Heller84} are prominent examples of 
``exceptional'' eigenstates  
violating  Eq.~(\ref{eq-Snirelman}). 
Choosing, e.g., a Gaussian weight of width $\Delta E$ in the energy average
in Eq.~(\ref{eq-correl_function}),
one can express $K_{ab}(T)$ for $T > \Delta T/2$ as
\begin{widetext}
\begin{eqnarray} \label{eq-exact_K_ab}
\nonumber
K_{ab} ( T ) 
&  = &  
  \frac{1}{ \langle d (E) \rangle_{\Delta E}} \frac{1}{\sqrt{2\pi \Delta E^2}}
   \sum_{n,m} \langle n | \hat{a} | n \rangle  \langle m | \hat{b} | m \rangle
     e^{ - \frac{\I}{\hbar} (E_m - E_n )T}   \, h ( E_m - E_n)
\\
& &      \times \, \exp \left( -  \frac{(E_n + E_m - 2 E)^2}{8 \Delta E^2} \right)
\,.
\end{eqnarray}
\end{widetext}
Since all functions of $E_n$ and $E_m$ vary noticeably on the scale
$\Delta E \gg \langle d(E) \rangle^{-1}_{\Delta E}$, 
the eigenstates not belonging to the ``set of density one'',
such as scars, have a negligible contribution 
to the sum  in Eq.~(\ref{eq-exact_K_ab}). 
One can then replace  
$\langle n | \hat{a} | n \rangle  \langle m | \hat{b} | m \rangle$
by the product
$\overline{a ({\bf x})} \; \overline{b ({\bf x})}$
and move this factor out of the sum.
This yields Eq.~(\ref{eq-leadingterm}), which is therefore  a direct consequence
of Snirelman's theorem.

To obtain information on matrix element fluctuations,
one thus needs to study the semiclassical corrections (next term in power of $\hbar$)
to the leading behavior  (\ref{eq-leadingterm}) of the form factor. 
Let us define
\begin{equation}
\hat{a}^\prime = \hat{a} - \overline{a ({\bf x})} \hat{1} \;\;\mbox{ , }\;\;
\hat{b}^\prime = \hat{b} - \overline{b ({\bf x})} \hat{1} \, ,
\end{equation}
so that the associated Weyl symbols 
$a^\prime ({\bf x})$ and $b^\prime( {\bf x})$  
average to zero. Then the form factor $K_{ab}(T)$ is related to 
$K_{a^\prime b^\prime} (T)$ by the formula
\begin{eqnarray} \label{eq-decomposition}
K_{ab} (T)  & - & 
  \overline{a ({\bf x})}\; \overline{b ({\bf x})} \, K(T) 
   =  \\
& \, & K_{a^\prime b^\prime} (T) 
    +  \overline{a ({\bf x})} \, K_{1 b^\prime} (T) +  \overline{b ({\bf x})}  \, 
      K_{a^\prime 1} (T)\,. \nonumber
\end{eqnarray}
Comparing with Eq.~(\ref{eq-leadingterm}), one sees that the r.h.s. of 
Eq.~(\ref{eq-decomposition}) vanishes as $\hbar \to 0$.
The purpose of the two next subsections is to estimate  
the first  term of the r.h.s., which turns out to be 
 proportional to $1/T_H = {\mathcal{O}} (\hbar^{f-1})$. 
We start with the diagonal contribution of pairs of
identical orbits (modulo time--reversal)
to $K_{a'b'}(T)$ and then include 
the pairs of correlated orbits $(\gamma , \gamma^p)$ 
studied in Section~\ref{sec:higher_dimensional}.
We restrict our derivation to the case of observables
with vanishing mean, i.e.,
$\overline{a ({\bf x})} = \overline{b ({\bf x})} =0$
so that $a = a^\prime$ and $b = b^\prime$.
Therefore we shall not be concerned further in this paper
with the second and third terms in Eq.~(\ref{eq-decomposition}).

\subsection{Correction term within the diagonal approximation}

Let us first consider the semiclassical correction 
to the leading term (\ref{eq-leadingterm})
within the  diagonal approximation.
This correction has been already studied 
in Refs.~\cite{Eckhardt95,EFKAMM95,Eckhardt97,Eckhardt00}. 
However, we will argue below that the
results of Refs.~\cite{Eckhardt95,EFKAMM95,Eckhardt97,Eckhardt00} 
can  only be applied to observables 
$a({\bf x})$ or $b({\bf x})$ independent of 
the momentum ${\bf p}$. We treat here 
the more general case of  smooth observables 
$a$ and ${b}$ depending on both the position ${\bf q}$ and  the momentum ${\bf p}$,  
by following the lines of Subsection IIC of Ref.~\cite{EFKAMM95}.

Retaining only the contribution of those pairs obtained by pairing
each orbit with itself or with its
time--reversed version
in the double sum (\ref{gen_K}), the semiclassical form factor can be written as
\begin{widetext}
\begin{equation} \label{eq-diag_approx_K_gen}
K_{a b}^{\text{(1)}} (T)
 =  
\frac{1}{T_H} 
 \left\langle 
   \int_0^{T_\gamma} \frac{{\rm d} t}{T_\gamma}  \, a ({\bf x}^\gamma_t ) \,  
     \left( \int_0^{T_\gamma} {\rm d} t'' \, b ({\bf x}^\gamma_{t''} )
      +    
     \int_0^{T_\gamma} {\rm d} t'' \, b ( \T {\bf x}^\gamma_{t''} ) 
     \right)
 \right\rangle_{\text{po},T}\, .
\end{equation}
\end{widetext}
Substituting $t'=t''-t$ and using the periodicity of $\gamma$
yields
\begin{equation} \label{eq-K_diag2}
K_{a b}^{\text{(1)}} (T)
 =  
\frac{2}{T_H} \int_0^{T} {\rm d} t\,
 \left\langle  
    \int_0^{T_\gamma} \frac{{\rm d} t'}{T_\gamma} \, 
   a ({\bf x}^\gamma_{t} ) 
    b^S ({\bf x}^\gamma_{t+t'} )
 \right\rangle_{\text{po},T}
\end{equation}
with $b^S({\bf x})$ as given in Eq.~(\ref{def:correl_ab}).

We now assume that the classical dynamics is sufficiently chaotic so that 
the time--correlation function (\ref{def:correl_ab}) of the classical 
observables $a({\bf x})$ and $b^S({\bf x})$ 
decays faster than $1/t$ to zero.
In strongly chaotic systems all 
classical correlation functions of smooth observables
decay exponentially, as a 
result of a gap in the spectrum of the resonances of the Frobenius--Perron operator 
(all resonances but the one corresponding to the Liouville measure are contained 
inside a circle of radius
strictly smaller than unity)~\cite{Gaspard98}. The mixing property makes sure
that the time--correlation function 
$C_{ab}^{S}(t)$ tends to zero in the large-$t$ limit, but is still not strong enough for
our purpose: it does 
{\it not} imply that $C_{ab}^{S}(t)$ can be integrated from $0$ to $\infty$. 
 
Applying  the sum rule (\ref{sum_rule}) to Eq.~(\ref{eq-K_diag2}) gives 
\begin{equation} \label{eq-diag_K}
K_{a b}^{\text{(1)}} ( T) 
 \simeq
\frac{2}{T_H} 
 \int_0^\infty {\rm d} t' \, C_{ab}^{S}(t') \;
\end{equation}
in the limit $\hbar \to 0$ with $\tau=T/T_H$ fixed.
If $a ( {\bf x})$ or $b ( {\bf x})$ is a function of the position ${\bf q}$ only,
then $a^S ({\bf x} ) = a ( {\bf x} )$ or 
$b^S ({\bf x} ) = b ( {\bf x} )$, respectively. As a result,
$C_{a b}^S (t) = C_{a b} (t)$. In such a case, Eq.~(\ref{eq-diag_K}) coincides
with the result of Refs.~\cite{Eckhardt95} and ~\cite{EFKAMM95}. In the opposite case,
the mean $C_{a b}^S (t)$ of the correlation function of
$a({\bf x})$ and $b ( {\bf x})$ and the correlation function of $a( {\bf x})$ and 
$b ( {\cal T} {\bf x} )$
must be considered.  

As noted in Ref.~\cite{EFKAMM95}, some chaotic systems which fulfill
the mixing property, such as the symmetric stadium billiard, 
exhibit algebraic decays of correlations in $1/t$.
Then the integral in Eq. (\ref{eq-diag_K}) diverges 
and the form factor $K_{ab}^{(1)}$ is of order $T_H^{-1} \ln T_H$ instead of  
$T_H^{-1}$.

Let us also mention that, for chaotic dynamics, the integrals
$\int_0^T {\rm d} t \, a ({\bf x}_t)$ and $\int_0^T {\rm d} t \, b^S({\bf x}_t )$ 
of the observables $a({\bf x} )$ and $b^S( {\bf x})$ along pieces of (nonperiodic)
trajectories of time $T$, thought 
as function of the initial point ${\bf x}$, may be often
considered as Gaussian random variables 
with respect to the Liouville measure for large $T$'s~\cite{Gaspard98}.
These random variables have a system--specific
covariance $2 T \int_0^\infty {\rm d} t \, C_{ab}^S (t)$,
which is thus also 
related to the fluctuations of the diagonal matrix elements of $\hat{a}$
and $\hat{b}$ as given by $K_{ab}(T)$. 

\subsection{Contribution of the partner orbits}

The contribution of the 
partner orbits to the semiclassical form factor (\ref{gen_K}) is obtained by inserting 
the product $ A_\gamma\,B_{{\gamma}^p}$
of the integrals  of the classical
observable $a ( {\bf x})$  along $\gamma$ and
of the observable  $b ( {\bf x})$ along $\gamma^p$
in front of the exponential in Eq.~(\ref{eq-spect_K_2}).
The forthcoming calculation is
simplified by noting that if ${\gamma}^p$ is a partner orbit of $\gamma$, then
its time--reversed version ${\gamma}^{p,i}$ is also a partner orbit of $\gamma$, with the same
action. This is because 
if one  exchanges the role of parts ${\cal R}$ and ${\cal L}$ in the definition~(\ref{partner_coords}),
the corresponding partner orbit is just ${\gamma}^{p,i}$.
As a result, one may  equivalently insert
$A_\gamma (B_{{\gamma}^p} + B_{{\gamma}^{p,i} } )/2$
in front  of the exponential in Eq.~(\ref{eq-spect_K_2}), instead of $ A_\gamma\,B_{{\gamma}^p}$. 
The mean $(B_{{\gamma}^p} + B_{{\gamma}^{p,i} } )/2$ is 
the integral  of the symmetrized observable
$b^S ( {\bf x})$ given by Eq.~(\ref{def:correl_ab}) along $\gamma^p$. It can be estimated by  
applying Eq.~(\ref{f_along_partner}) to $b^S({\bf x})$ and using $b^S({\bf x}) = b^S ( \T {\bf x})$
together with the periodicity of $\gamma$. This yields
\begin{equation} 
\frac{1}{2} \left( B_{{\gamma}^p } + B_{ {\gamma}^{p,i} } \right) 
    \simeq   
     \int_0^{T_\gamma} \frac{{\rm d} t'' }{ T_{\gamma}}\,b^S ( {\bf x}_{t''}^\gamma )\;
\end{equation}  
and reflects the fact that the two orbits $\gamma$ and $\gamma^i$
explore almost the same phase--space regions
as the two partner orbits $\gamma^p$ and
$\gamma^{p,i}$.
Hence the generalization of Eq.~(\ref{K2_partners}) reads
\begin{widetext}
\begin{eqnarray}
\label{K2gen_partners}
\nonumber
 K_{ab}^{(2)} (\tau)  = \tau & & \left\langle
 \int\limits_{-S_{\text{max}}(E)}^{S_{\text{max}}(E)}
 {\rm d}S_1 \dots {\rm d}S_{f-1} \, 
 \left\langle \frac{{\rm d}^{f-1} N_\gamma(\{S_i\})}
                   {{\rm d}S_1 \dots {\rm d}S_{f-1}}
 \int_0^{T_\gamma} \frac{ {\rm d} t'}{T_\gamma} 
    \int_0^{T_\gamma} \frac{ {\rm d} t''}{T_\gamma}  
   \,a ({\bf x}_{t'}^\gamma ) b^S ({\bf x}_{t' + t''}^\gamma )
 \right\rangle_{{\rm po},\tau T_H} 
  \right.
\\
& & 
  \left.
  \times  \exp \left( {\rm i}
  \sum\limits_{i=1}^{f-1} \frac{S_i}{\hbar} \right) \right\rangle_{\Delta E} \; .
\end{eqnarray}
\end{widetext}
By using Eq.~(\ref{dN_flow}) and substituting
$t' \to t ''' = t' - t$ before applying  
the sum rule (\ref{sum_rule}), one finds
 that the
leading contribution $\rho^{\text{lead}}$ to the density  in Eq.~(\ref{result:density}) yields
\begin{widetext}
\begin{eqnarray} \label{eq_N_gen_flux}
\nonumber
& \displaystyle
\left\langle \frac{{\rm d}^{f-1} N_\gamma(\{S_i\})}
                   {{\rm d}S_1 \dots {\rm d}S_{f-1}}
 \int_0^{T_\gamma} \frac{ {\rm d} t'}{T_\gamma} 
    \int_0^{T_\gamma} \frac{ {\rm d} t''}{T_\gamma}  
   \,a ({\bf x}_{t'}^\gamma ) b^S ({\bf x}_{t' + t''}^\gamma )
\right\rangle_{{\rm po},T}^{\text{(lead)}} \qquad \qquad \qquad \qquad 
\\
& \displaystyle
 \approx  \frac{2^{f-1}}{\Omega} \sum_{j=1}^{f-1} \prod_{i \not= j}^{f-1} 
  \ln \left( \frac{S_{\text{cl}} c^2}{|S_i|} \right)\
    \int_0^T {\rm d} t''' \int_0^T {\rm d} t'' \, \overline{\chi_j ({\bf x}) \,
       a( {\bf x}_{t'''} ) b^S ( {\bf x}_{t'''+t''} )} \;.
\end{eqnarray}
\end{widetext}
Employing ergodicity, the integral over $t'''$ can be approximated by a phase--space average and yields
$T\, \overline{\chi_j ( {\bf x})} \int_0^\infty \D t'' \,C_{ab}^S ( t'')$.
Thus inserting Eq.~(\ref{eq_N_gen_flux}) into Eq.~(\ref{K2gen_partners}) gives
$K_{ab}^{(2)} (\tau) =0$.
As in Section~\ref{sec:higher_dimensional}, the contribution 
to the form factor of $\rho^{\text{lead}}$ thus vanishes. 
By using Eq.~(\ref{dN_volume}), we obtain the contribution of the small correction term $\rho^{\text{(corr)}}$ in Eq.~(\ref{result:density}),
\begin{widetext}
\begin{eqnarray} \label{eq_N_gen_vol}
\nonumber
& \displaystyle
\left\langle \frac{{\rm d}^{f-1} N_\gamma(\{S_i\})}
                   {{\rm d}S_1 \dots {\rm d}S_{f-1}}
 \int_0^{T_\gamma} \frac{ {\rm d} t'}{T_\gamma} 
    \int_0^{T_\gamma} \frac{ {\rm d} t''}{T_\gamma}   
   \,a ({\bf x}_{t'}^\gamma ) b^S ({\bf x}_{t' + t''}^\gamma )
\right\rangle_{{\rm po},T}^{\text{(corr)}}  
\qquad \qquad
\\
& \displaystyle
 \approx 
  - 2^f \frac{1}{\Omega T} \prod_{i=1}^{f-1} 
   \ln \left( \frac{S_{\text{cl}} c^2}{|S_i|} \right)
    \int_0^T {\rm d} t''' \int_0^T {\rm d} t'' \,
       \overline{ a( {\bf x}_{t'''} ) b^S ( {\bf x}_{t'''+t''} )}       
\end{eqnarray}
\end{widetext}
since the dependence on $t_{\text{enc}} ( \{ s_i, u_i \} ; {\bf x} )$ in
Eq.~(\ref{dN_volume}) and Eq.~(\ref{result:density}) mutually cancels.
The average $\overline{ a( {\bf x}_{t'''} ) b^S ( {\bf x}_{t'''+t''} )}$ 
equals the correlation function $C_{ab}^S ( t'')$.
It follows  from Eq.~(\ref{result:K2}) that, as $\hbar \to 0$,
\begin{equation} \label{eq-corr_K_ab}
K_{ab}^{(2)} (\tau) \approx - 2 \tau \, 
\frac{1}{T_H} \int_0^\infty {\rm d} t \,C_{ab}^S (t) \; .
 \end{equation}
Remarkably, one obtains  for the leading off-diagonal contribution
the same result as for the diagonal approximation, with $ 2$ replaced by
$-2 \tau$ as in the spectral form factor.
In particular this means that the classical correlations
enter in exactly the same way via the correlation function
$C_{ab}^S(t)$. Assuming that 
only  identical orbits modulo time--reversal symmetry and pairs of 
partner orbits $(\gamma, \gamma^p)$ contribute to the semiclassical form factor 
(\ref{gen_K}) up to order 
$\tau^2$ included, this yields 
\begin{equation} \label{eq-gen_K_result}
K_{ab} (\tau) \approx \frac{1}{\tau \; T_H} \Bigl( K(\tau) + {\mathcal{O} } ( \tau^3) \Bigr)
\int_0^\infty {\rm d} t \,C_{ab}^S (t )  
\end{equation}
as announced in the introduction.
This result holds if the correlation function
$C_{ab}^S(t)$ decays  faster than $1/t$ as 
$t \to \infty$,
in order that  the upper integration limit  $T = \tau T_H$
may be replaced by $\infty$. It is valid 
for observables $\hat{a}$ and $\hat{b}$ such that
$\overline{a ({\bf x})} = \overline{b ({\bf x})} = 0$ only. 

If $\overline{a ({\bf x})} \not=0$ or 
$\overline{b ({\bf x})} \not=0$, 
 Eq.~(\ref{eq-decomposition}) must be used and  the second and third terms in the r.h.s. 
of this equation have to be estimated. 
Repeating the above calculation for these terms, one finds that both vanish in 
zeroth order in $\hbar$, thus being in accordance with  Snirelman's theorem. 
For instance, within the diagonal approximation, Eq.~(\ref{eq-diag_approx_K_gen}) gives 
$K_{a' 1}^{(1)}(T) 
= 2 T_H^{-1}  \langle \int_0^{T_\gamma} \D t\, a'( {\bf x}_t^\gamma ) \rangle_{\text{po},T}
\approx 2 \tau\,\overline{a' ( {\bf x})}$, which is zero since $\overline{a' ( {\bf x})} =0$.
The leading contribution in $\hbar$ is thus governed by the finite--time 
corrections to the sum rule (\ref{sum_rule}). 
Similarly, replacing $a$ by $a'$ and $b$ by $1$ in Eq.~(\ref{eq_N_gen_vol}), the second integral
in the second member becomes $T \,\overline{a'({\bf x})}$, which means that 
$K_{a' 1}^{(2)}(T) \approx 0$ up to higher--order corrections
in the sum rule. One concludes that our method 
does not allow us to estimate $K_{a' 1}$ as $\hbar \to 0$
beyond the leading order in $\hbar$. 
For systems with exponential decay of classical correlation functions,
it is not irreasonable to expect that the finite--time 
corrections to the sum rule (\ref{sum_rule}) are exponentially 
small in $T$. In such a case
$K_{a' 1}$ and $K_{1 b'}$ would be negligible with respect to $K_{a'b'}$, 
which is of order $\hbar^{f-1}$ by
Eq.~(\ref{eq-gen_K_result}).

\section{Summary and outlook}
In this work we presented a semiclassical evaluation of the
generalized form factor $K_{ab}(\tau)$ going beyond the diagonal
approximation. We first considered the spectral
form factor $K(\tau)=K_{11}(\tau)$ for systems with more than two
degrees of freedom, i.e., $f\geq2$. We proved that the leading
contribution due to pairs of periodic orbits with 
correlated actions is independent of $f$ in agreement with the
RMT prediction. An important step in our calculation was to  show the equivalence
between the two different approaches for counting partner orbits which were independently
developed in Ref.~\cite{Turek03} and Refs.~\cite{Spehner03,Heusler03}
for two--dimensional systems. Based on these
results for the spectral form factor 
we then investigated the
generalized form factor $K_{ab}(\tau)$.
In this case we were able to show
a universal dependence of $K_{ab}(\tau)$ on the rescaled time
$\tau$. Furthermore, we found that the contribution of the partner orbits
depends on the classical time--correlation function
$C_{ab}^{S}(t)$ in exactly the same way as in the diagonal approximation, 
see Eq.~(\ref{eq-gen_K_result}). 
An interesting open question is to prove (or disprove) that this is still the case
at higher orders in $\tau$ or even for arbitrary large $\tau > 1$. 
In such a case one could get rid of the error term 
${\cal O} (\tau^3)$ added to $K(\tau)$ in Eq.~(\ref{eq-gen_K_result}).

Our semiclassical treatment of the generalized form factor beyond
the diagonal approximation can in principle be extended 
to other physical observables 
containing matrix elements in chaotic systems. This
includes expressions where transition matrix elements
play a role~\cite{Wilk87,Mehlig98}
(e.g., dipole excitations in quantum dots~\cite{MR98}),
and linear response functions for 
mesoscopic systems~\cite{Ric00} with 
applications to transport, magnetism, or optical 
response. So far, nearly all semiclassical approaches to 
such quantities have been relying  on the diagonal approximation, 
as long as additional averages are involved. A notable exception is
the calculation of the weak localization correction to the 
conductance in Ref.~\cite{Ric02}, showing an important contribution
of the partner orbits. It would thus be of great interest to 
study the corrections to the diagonal approximation
in the various response functions appearing in mesoscopic
physics.

\vspace*{0.3cm}

{\bf Acknowledgments:}
We acknowledge support from the  {\rm Deutsche
Forschungsgemeinschaft} (Ri 681/5 and SFB/TR 12).
We are grateful to B.~Eckhardt and U.~Smilansky
for interesting discussions. SM also thanks P.~Braun,
F.~Haake, and S.~Heusler for close cooperation. 

\appendix*

\section{Transformation of the surface integral into a volume integral}
\label{app:proof}

In this appendix we prove the equality of the
two different approaches for counting the partner
orbits based on Eq.~(\ref{dN_flow})
and Eq.~(\ref{dN_volume}), respectively.
To this end we show that an equality of the
general structure 
\begin{equation}
\label{gen_flow_volume}
 \int\limits_0^T {\rm d} t  \,  
 \int\limits_{\cal V} {\rm d} V_{\delta \vec{y}} \;
 \frac{\varrho(\delta \vec{y} , t)}{t_{\cal V}(\delta \vec{y} , t)}
 \; = \; 
 \int\limits_0^T {\rm d} t \, \int\limits_{\partial {\cal V}_{\rm out}}
 {\rm d} \vec{A}_{\delta \vec{y}} \; \varrho(\delta \vec{y} , t) \; 
 \dot{\delta \vec{y}}(\delta \vec{y} , t) \, 
\end{equation}
holds under the conditions which are relevant for the statistics of the number
of partners. Here, $\delta \vec{y}$ is a vector in a multidimensional space, e.g.,
the $(2f-2)$-dimensional PSS. The volume element in this space is given by
${\rm d} V_{\delta \vec{y}} = {\rm d}^{f-1} u \, {\rm d}^{f-1} s$
while ${\rm d} \vec{A}_{\delta \vec{y}}$ characterizes
the surface element. The left hand side of Eq.~(\ref{gen_flow_volume})
thus contains an integral over any $(2f-2)$-dimensional 
volume ${\cal V}$ in the PSS.
Inside ${\cal V}$ we follow the time evolution of a density field
$\varrho(\delta \vec{y} , t)$; the corresponding velocity field is
denoted by
${\dot{\delta \vec{y}}}(\delta \vec{y} , t)$.
As Eq.~(\ref{gen_flow_volume}) is applied to the PSS following a
periodic orbit of length $T$, we can assume periodicity
such that $\varrho(\delta \vec{y} , t) = \varrho(\delta \vec{y} , t+T)$
and $\dot{\delta \vec{y}}(\delta \vec{y} , t) =
 \dot{\delta \vec{y}}(\delta \vec{y} , t+T)$.
Due to current conservation the density
is constant along the flow, i.e.,
$\varrho(\delta \vec{y},0) = \varrho(\delta \vec{y}_t , t)$ or
$\dot{\varrho}(\delta \vec{y}_t,t) = 0$. The
time $t_{\cal V}(\delta \vec{y} , t)$ in Eq.~(\ref{gen_flow_volume})
is defined as the
total time a point spends in the volume ${\cal V}$ if
it starts at time $t$ at position $\delta \vec{y}$ and
moves until time $t+T$. If the volume
${\cal V}$ is chosen to coincide with the hypercube
${\cal C}$ defining the encounter region, see 
Subsection~\ref{subsec:encounter}, then $t_{\cal V}$ is
approximately equal to the time $t_{\text{enc}}$,
Eq.~(\ref{t_enc}).
The surface of ${\cal V}$ is decomposed as
$\partial {\cal V} = \partial {\cal V}_{\rm in} +
 \partial {\cal V}_{\rm out}$. Here,
$\partial {\cal V}_{\rm in/out}$ stands for that part
of the total surface through which the flux defined
by $\varrho$ and $\dot{\delta \vec{y}}$
enters or leaves ${\cal V}$ in the long--time limit, respectively.
More precisely speaking, the total flux between
time $0$ and $T$ through
any piece of $\partial {\cal V}_{\rm out}$
must be positive.

For the proof of relation (\ref{gen_flow_volume})
let us first consider the case where the total
density $\varrho(\delta \vec{y} , t)$ is given by a single
point starting at $\delta \vec{y}_0$, i.e.,
$\varrho_{1}(\delta \vec{y},t) = \delta ( \delta \vec{y} - \delta \vec{y}_t )$.
Then the time $t_{\cal V}$ is given as
\begin{eqnarray}
\label{t_v}
        t_{\cal V}(\delta \vec{y}_t , t) 
         & = &  \int\limits_t^{t+T} {\rm d} t' \; \Theta_{\cal V}(\delta \vec{y}_{t'})
        = \int\limits_0^T {\rm d} t' \; \int\limits_{\cal V} {\rm d}
        V_{\delta \vec{y\,'}} \; \varrho_{1}
           (\delta \vec{y\,}',t') \nonumber \\ 
        & = &  t_{\cal V}(\delta \vec{y}_0 , 0) \; ,
\end{eqnarray}
where $\Theta_{\cal V}(\delta \vec{y})$ equals $1$ if $\delta \vec{y} \in {\cal V}$
and zero otherwise. In Eq.~(\ref{t_v}) we made use
of the periodicity of the motion. We then obtain for the left
hand side of Eq.~(\ref{gen_flow_volume})
\begin{eqnarray}
        \int\limits_0^T {\rm d} t & \, & \int\limits_{\cal V} {\rm d}
           V_{\delta \vec{y}}
        \; \frac{\varrho_{1}(\delta \vec{y},t)}
        {t_{\cal V}(\delta \vec{y} , t)}
         =  \\
        & \, &
        \frac{1}{t_{\cal V}(\delta \vec{y}_0 , 0)} \int\limits_0^T {\rm d} t \,
        \int\limits_{\cal V} {\rm d} V_{\delta \vec{y}}
        \; \varrho_{1}(\delta\vec{y},t) = 1 \, . \nonumber
\end{eqnarray}
In close analogy we thus find that if the single point density is replaced by
$\varrho(\delta \vec{y} , t) = \sum_{i} 
 \varrho_{i} (\delta \vec{y} , t)$
which represents an arbitrary number $n$ of points given
by their initial conditions then the left hand side
of Eq.~(\ref{gen_flow_volume}) just gives the total number
of particles $n$ that pass ${\cal V}$ during one period.
But this is exactly what the right hand side of 
Eq.~(\ref{gen_flow_volume}) gives. It just measures the
outgoing flux through the surface of $\cal V$ between time
$0$ and $T$ which also yields the total number
of particles $n$ because the particle number is conserved.

Finally we also note that the density $\varrho(\delta \vec{y} , t)$ 
is not restricted to a sum of $\delta$ functions. Each of these
$\delta$ functions can also be multiplied with
any function $g(\delta \vec{y} , t)$ that is constant when
following the flow within ${\cal V}$,
i.e., $g(\delta \vec{y}_0 , 0) = g(\delta \vec{y}_t , t)$.
In the context of Subsection~\ref{subsec:statistics_of},
$g$ could, for example, be any function of the action difference as
in Eqs.~(\ref{dN_flow}) and~(\ref{dN_volume}).
In this case the density $\varrho$ entering Eq.~(\ref{gen_flow_volume})
can be considered as a weighted density $\varrho = \rho \, g$.

If all local unstable growth rates $\chi_k({\bf x})$
are non-negative one can directly identify $t_{\cal V} = t_{\text{enc}}$
and thus the equality (\ref{gen_flow_volume}) means
that Eq.~(\ref{dN_flow}) exactly equals Eq.~(\ref{dN_volume}). On the
other hand, if these local unstable growth rates assume negative
values in certain areas of the phase space then this implies that
the unstable components of a displacement vector can also
decrease on {\em short} time scales. This would lead to a
multiple entry of the same point into the
'encounter region' characterized by ${\cal V} = {\cal C}$.
In this case the relation (\ref{gen_flow_volume})
means that Eq.~(\ref{dN_volume}) is asymptotically equal to
Eq.~(\ref{dN_flow}) as the length $t_{\text{enc}}$ becomes large
so that $|t_{\cal V} - t_{\text{enc}}| \ll t_{\text{enc}}$ or
similarly $t_{\cal V} \simeq t_{\text{enc}}$.


\end{document}